\documentclass[fleqn,usenatbib]{mnras}

\usepackage{newtxtext,newtxmath}

\usepackage[T1]{fontenc}
\usepackage{ae,aecompl}

\usepackage{graphicx}	
\usepackage{amsmath}	
\usepackage{mathtools}
\usepackage{hyperref}
\usepackage{xspace}	
\usepackage{euscript}

\usepackage{xcolor}	
\usepackage{pdflscape}
\usepackage{afterpage}
\usepackage{placeins}
\usepackage{bm}
\usepackage{ dsfont }
\usepackage{listings}
\usepackage{soul}
\usepackage{caption}
\usepackage{subcaption}
\usepackage{soul}

\usepackage{amsthm}

\usepackage{algorithm}
\usepackage[noend]{algpseudocode}

\definecolor{codegreen}{rgb}{0,0.6,0}
\definecolor{codegray}{rgb}{0.5,0.5,0.5}
\definecolor{codepurple}{rgb}{0.58,0,0.82}
\definecolor{backcolour}{rgb}{0.95,0.95,0.92}

\lstdefinestyle{python}{
    backgroundcolor=\color{backcolour},   
    commentstyle=\color{codegreen},
    keywordstyle=\color{magenta},
    numberstyle=\tiny\color{codegray},
    stringstyle=\color{codepurple},
    basicstyle=\ttfamily\footnotesize,
    breakatwhitespace=false,         
    breaklines=true,                 
    captionpos=b,                    
    keepspaces=true,                 
    numbers=left,                    
    numbersep=5pt,                  
    showspaces=false,                
    showstringspaces=false,
    showtabs=false,                  
    tabsize=2,
    upquote=true
}

\newcommand{\gaia}{{\it Gaia}\xspace}

\newcommand{\prob}{\mathrm{P}}
\defcitealias{Paper1}{Paper~I}
\defcitealias{Paper2}{Paper~II}
\defcitealias{Paper5}{Paper~V}
\defcitealias{Paper3}{Paper~III}
\defcitealias{Lindegren2018}{L18}

\title[The \gaia Astrometry Spread Function.]{Completeness of the \textit{Gaia}-verse IV: The Astrometry Spread Function of \textit{Gaia} DR2}

\author[A. Everall, D. Boubert, S. E. Koposov, L. Smith and B. Holl]{
	Andrew Everall$^{1}$\thanks{E-mail: asfe2@cam.ac.uk},
	Douglas Boubert$^{2}$, 
	Sergey E. Koposov$^{3,1,4}$, 
	Leigh Smith$^{1}$
	and Berry Holl$^{5,6}$
	\\
	$^{1}$Institute of Astronomy, University of Cambridge, Madingley Road, Cambridge CB3 0HA, UK\\
	$^{2}$Magdalen College, University of Oxford, High Street, Oxford OX1 4AU, UK\\
    $^{3}$Institute for Astronomy, University of Edinburgh, Royal Observatory, Blackford Hill, Edinburgh EH9 3HJ, UK\\
    $^{4}$McWilliams Center for Cosmology, Carnegie Mellon University, 5000 Forbes Ave, 15213\\
	$^{5}$ Department of Astronomy, University of Geneva, Ch. des Maillettes 51, CH-1290 Versoix, Switzerland \\
    $^{6}$ Department of Astronomy, University of Geneva, Ch. d’Ecogia 16, CH-1290 Versoix, Switzerland \\
}

\date{Accepted XXX. Received YYY; in original form ZZZ}

\pubyear{}

\begin{document}
\label{firstpage}
\pagerange{\pageref{firstpage}--\pageref{lastpage}}
\maketitle

\begin{abstract}
\gaia DR2 published  positions, parallaxes and proper motions for an unprecedented 1,331,909,727 sources, revolutionising the field of Galactic dynamics. We complement this data with the Astrometry Spread Function (ASF), the expected uncertainty in the measured positions, proper motions and parallax for a non-accelerating point source. The ASF is a Gaussian function for which we construct the 5D astrometric covariance matrix as a function of position on the sky and apparent magnitude using the \gaia DR2 scanning law and demonstrate excellent agreement with the observed data. This can be used to answer the question `What astrometric covariance would \gaia have published if my star was a non-accelerating point source?'.

The ASF will enable characterisation of binary systems, exoplanet orbits, astrometric microlensing events and extended sources which add an excess astrometric noise to the expected astrometry uncertainty. By using the ASF to estimate the unit weight error (UWE) of \gaia DR2 sources, we demonstrate that the ASF indeed provides a direct probe of the excess source noise.

 We use the ASF to estimate the contribution to the selection function of the \gaia astrometric sample from a cut on \textsc{astrometric\_sigma5d\_max} showing high completeness for $G<20$ dropping to $<1\%$ in underscanned regions of the sky for $G=21$.

We have added an ASF module to the \textsc{Python} package \textsc{scanninglaw} (\url{https://github.com/gaiaverse/scanninglaw}) through which users can access the ASF.
\end{abstract}




\section{Introduction}


\gaia has initiated an era of large scale Milky Way dynamical modelling by providing 5D astrometry (position, proper motion and parallax) for more than 1.3 billion stars \citep{Prusti2016,Brown2018, Lindegren2018}. The \gaia satellite measures source positions at multiple epochs over the mission lifetime. These epoch astrometry measurements are the inputs of the Astrometric Global Iterative Solution \citep[AGIS][]{Lindegren2012} which iteratively solves for the spacecraft attitude, geometric calibration of the instrument, global parameters, and 5D astrometry of each source: the right ascension and declination ($\alpha_0$, $\delta_0$), the proper motions ($\mu_\alpha$, $\mu_\delta$) and the parallax ($\varpi$). Alongside the source astrometry, \gaia also publishes the 5D astrometric measurement covariance and various statistics of the astrometric solution for all sources which meet the quality cuts. The 5-parameter astrometric model of AGIS assumes sources are point-like with apparent non-accelerating uniform motion relative to the solar system barycenter which we will refer to as `simple point sources'.

Both resolved and unresolved binary stars accelerate due to their orbits around the common center of mass which shifts the centroid off a uniform motion trajectory. For example, given the full epoch astrometry, it is expected that \gaia can characterise the orbits of stars with brown dwarf companions out to 10 pc and black hole companions out to more than 1 kpc when considering tight constraints of the uncertainty on the mass function $M^3_2 M^{-2}_{\hbox{tot}}$ \citep{Andrews2019}. When one is interested in (the less constraining) orbital parameter recovery with $\sim10$\% precision, \gaia might detect a staggering 20~k brown dwarfs around FGK-stars out to many tens up to a few hundreds of pc for the longer period objects (100-3000~d), which could reach even 50~k out to several hundred pc when one is only interested in the detection of BD candidates (e.g. for follow-up studies) \citep{HollBd}, with black hole companions being detectable out to several kpc.


Similarly exoplanet orbits pull their host stars away from uniform motion although with a much smaller amplitude due to the lower companion mass. From simulations it is expected that \gaia is capable of detecting 21,000 long-period, 1-15 Jupiter mass planets during the 5 year mission \citep{Perryman2014}, more than 4 times the number of currently known exoplanets. \citet{Ranalli2018} have further demonstrated that the 5 year \gaia mission will be able to find Jupiter-mass planets on 3 au orbits around $1M_\odot$ stars out to 39 pc and Neptune-mass planets out to 1.9 pc. Not only will the presence of planets be detectable but it is expected that $\sim500$ planets around M-dwarfs will receive mass constraints purely from \gaia astrometry \citep{Casertano2008, Sozzetti2014}.  

Microlensing occurs when the light from a background source is gravitationally lensed by a foreground lensing star causing a shift in the apparent position of the source, detectable by high precision astrometric surveys \citep{Miralda1996}. The deflection can be used as a direct measurement of the lens mass as demonstrated by \citet{Kains2017} using HST observations. A signficant amount of work has gone towards predicting microlensing events in using \gaia proper motions \citep{Kluter2018, Bramich2018, McGill2019} with $528$ events expected in the extended \gaia mission $\sim39\%$ of which pass astrometry quality cuts \citep{McGill2020}. For a small number of these events \gaia will be able to determine the lens mass to $<30\%$ uncertainty \citep{Kluter2020m}.

Extended sources such as galaxies will have a reduced astrometric precision from each \gaia observation due to the increased spread of flux. \gaia scans a source in many different directions over the mission lifetime from which the source shape can be reconstructed \citep{Harrison2011}. With the \gaia epoch astrometry for the 5 year mission, \gaia will be able to distinguish between elliptical and spiral/irregular galaxies with $\sim83\%$ accuracy \citep{KroneMartins2013}. These classifications would be incredibly valuable for galaxy morphology studies.

The \gaia epoch astrometry will be first released in DR4, several years from now. However, a (very) condensed form of this large amount of information is stored in the summary statistics of the astrometric solution currently published in \gaia DR2 and updated in EDR3. Binary stars, exoplanet hosts, microlensing events and extended sources will induce an excess noise in the astrometric solution as they are not well described by simple point sources. This excess noise has been modelled for binaries \citep{Wielen1997, Penoyre2020} and already \citet{Belokurov2020} has found many binaries in \gaia DR2 using the renormalised unit weight error statistic, RUWE, that is the re-normalised square root of the reduced $\chi^2$ statistic of the astrometric solution.

RUWE is a 1D summary statistic of the residuals of the 
5-parameter astrometric solution of a source relative to the \gaia inertial rest frame. But we can glean even more information on the excess noise from the 5D uncertainty of the astrometric solution. 
The uncertainty in the 5D astrometric solution for a source in \gaia can be expressed as the convolution of \gaia's astrometric measurement uncertainty expected for a simple point source and excess noise. We term \gaia's expected astrometric measurement uncertainty the Astrometry Spread Function (ASF) defined as the probability of measuring a simple point source to have astrometry $\mathbf{r}' \in \mathbb{R}^{5}$ given the true source astrometry $\mathbf{r} \in \mathbb{R}^{5}$ and apparent magnitude $G$ 
\begin{equation}
    \mathrm{ASF}(\mathbf{r}') = \mathrm{P}(\mathbf{r}'\,|\,\mathbf{r}, G).
\end{equation}
The excess noise will be driven by un-modelled source characteristics such as binary motion, exoplanet host motion, microlensing or extended source flux as well as any calibration noise which is not accounted for in the ASF. In this work, we'll assume that all significant calibration effects are included in the ASF such that the excess noise is dominated by un-modelled source characteristics. However this assumption breaks down in some regimes, particularly for bright sources in crowded regions where CCD saturation becomes a significant issue. Possible un-accounted calibration effects should be considered when using \gaia astrometry to search for excess noise due to genuine un-modelled source characteristics.

Since the astrometric solution is evaluated using least squares regression, the ASF will be Gaussian distributed
\begin{equation}
    \mathrm{ASF}(\mathbf{r}') = \mathcal{N}(\mathbf{r}'\,;\,\mathbf{r}, \Sigma(l,b,G))
\end{equation}
where $\Sigma(l,b,G) \in \mathbb{R}^{5\times5}$ is the expected covariance for a simple point source with position $l,b$ and apparent magnitude $G$ as measured by \gaia. The astrometric calibration is also a function of source colour which was either estimated from $G_\mathrm{BP}-G_\mathrm{RP}$ or added as a sixth parameter of the astrometric solution, \textsc{astrometric\_pseudo\_colour}. As colours are only published for a subset of the \gaia catalogue and the astrometric correlation coefficients for pseudo-colour are not published in DR2, we neglect colour dependence of the astrometric solution in this work. For EDR3, all pseudo-colour correlation coefficients are published and it will be worth considering how this impacts the ASF.


Given the ASF and published astrometric 5-parameter model uncertainties we can reconstruct the 5D excess noise and use it to characterise binary systems, exoplanet orbits, microlensing events and extended sources in \gaia without requiring the epoch astrometry. The focus of this paper is to construct the ASF for \gaia DR2.


This builds on analysis of the scanning law from \citet{Paper1} and \citet{Paper3} and will be used, in conjunction with the results of \citet{Paper2} to determine the selection function for the subsample of \gaia DR2 with published 5D astrometry.

In Section~\ref{sec:data} we provide a whistle-stop tour of the \gaia spacecraft, scanning law and how this translates to constraints on the position, proper motion and parallax of sources. This paper is focused on \gaia DR2 for which we estimate the ASF, although we note that the method will be directly applicable to \gaia EDR3. The method for constructing the ASF of \gaia is derived in Section~\ref{sec:method} and the results compared with the astrometry sample are shown in Section~\ref{sec:results}. We will also use the ASF for an alternative derivation of the Unit Weight Error demonstrating the applicability of the method in Section~\ref{sec:uwe}. 

As a secondary motivation, the \gaia DR2 5D astrometry sample is selected from the full catalogue with a cut on the parameter \textsc{astrometric\_sigma5d\_max} which is a function of the astrometric covariance matrix. In predicting the astrometric covariance for simple point sources, we can also estimate the contribution from this cut to the astrometric selection function which we will present in Section~\ref{sec:sigma5dmax}.

Finally we will discuss applications of the ASF in Section~\ref{sec:discussion} and provide instructions for accessing the data in Section~\ref{sec:code} before concluding.

\section{Astrometry with Gaia}
\label{sec:data}

\begin{figure}
    \centering
    \begin{subfigure}[t]{.495\textwidth}
    \centering
    \includegraphics[width=0.85\textwidth]{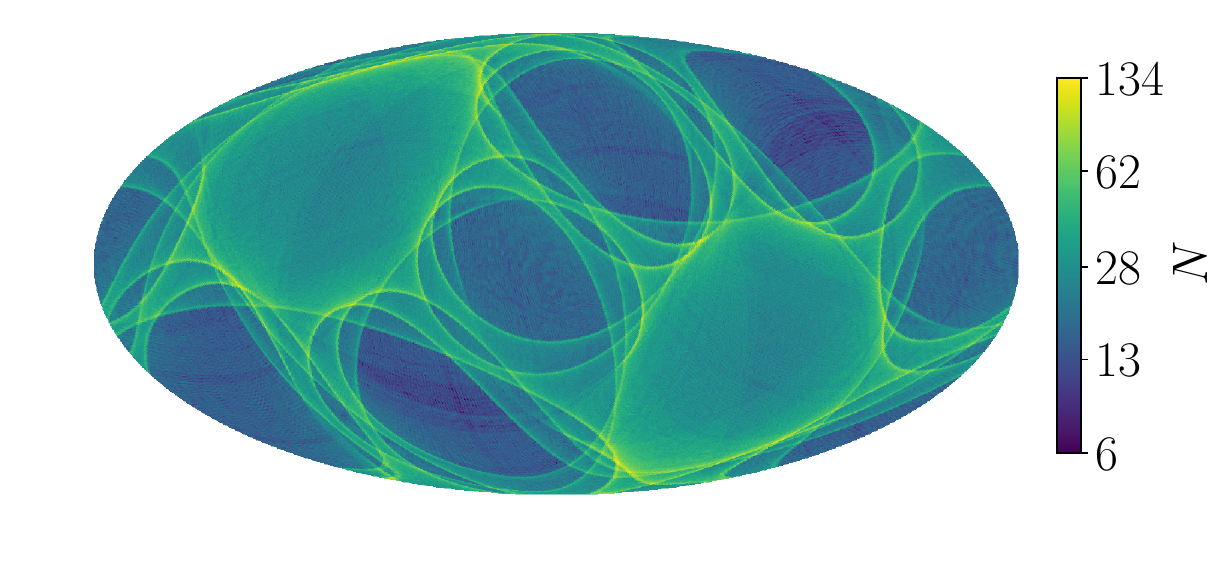} 
    \vspace*{-5mm}
    \caption{Regions which received more scans in DR2 (light yellow) produce tighter constraints on the astrometry whereas poorly scanned regions, including the Galactic bulge will have weaker inference.} \label{fig:slstats1}
    \end{subfigure}
    \begin{subfigure}[t]{.495\textwidth}
    \centering
    \includegraphics[width=0.85\textwidth]{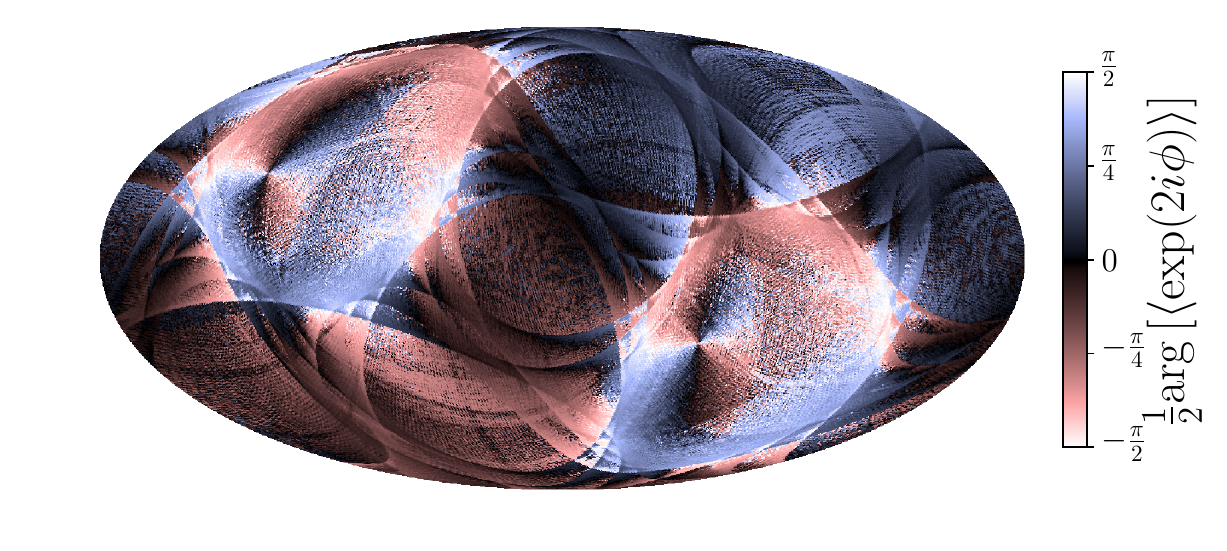} 
    \vspace*{-5mm}
    \caption{\gaia produces stronger measurements in the AL direction therefore the astrometry will be better constrained in the mean scan direction. Areas with more Equatorial polar scans (black) will constrain declination whilst lateral scans (white) constrain right ascension.} \label{fig:slstats2}
    \end{subfigure}
    \begin{subfigure}[t]{.495\textwidth}
    \centering
    \includegraphics[width=0.85\textwidth]{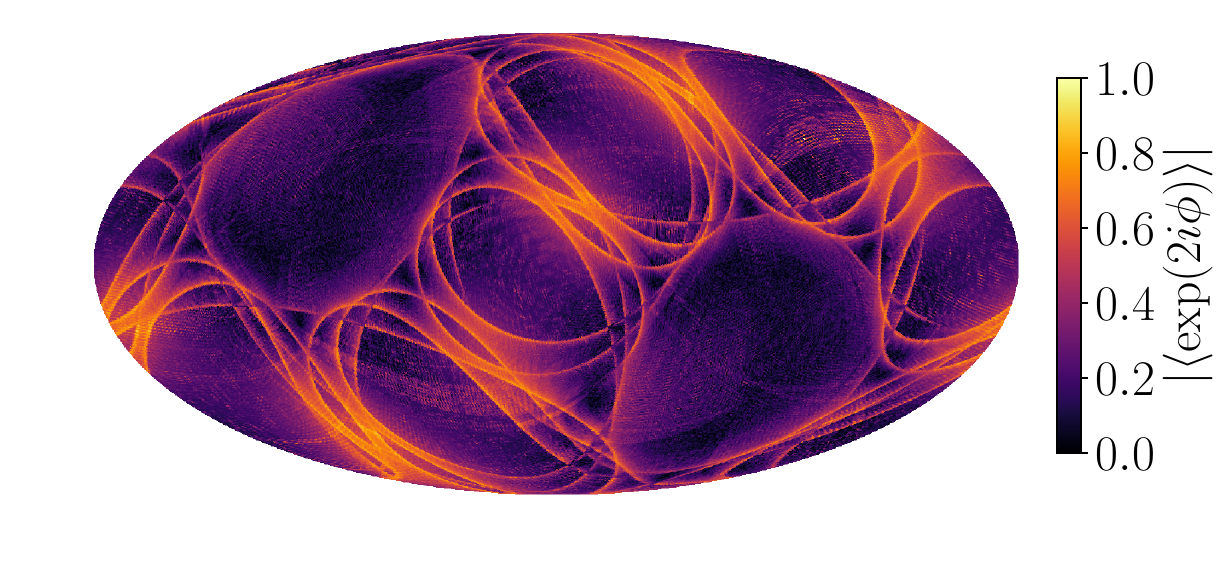} 
    \vspace*{-5mm}
    \caption{The significance in the difference in directional constraints is reflected by the clustering of scan directions. Heavily clustered scan directions (light yellow) will produce a much stronger constraint in the mean scan direction than the perpendicular direction.} \label{fig:slstats3}
    \end{subfigure}
    \begin{subfigure}[t]{.495\textwidth}
    \centering
    \includegraphics[width=0.85\textwidth]{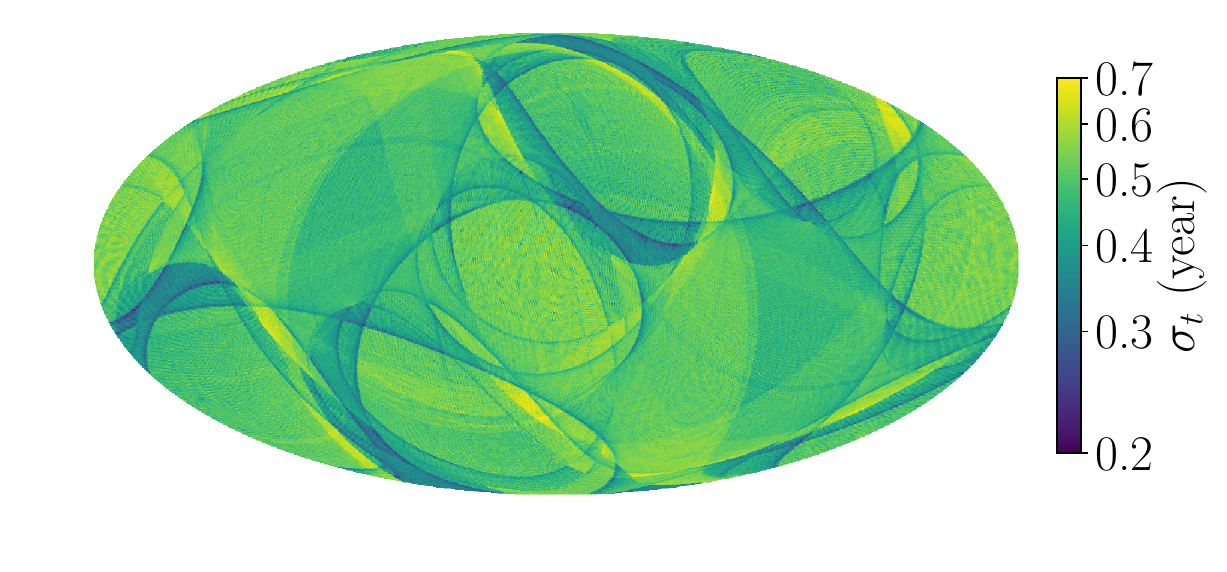} 
    \vspace*{-5mm}
    \caption{The spread of scan times, shown here by the standard deviation of times at which a position on the sky was observed, determines how well the proper motion can be estimated. A small spread in observation times 
    (dark blue) will provide weaker proper motion constraints.} \label{fig:slstats4}
    \end{subfigure}
    \begin{subfigure}[t]{.495\textwidth}
    \centering
    \includegraphics[width=0.85\textwidth]{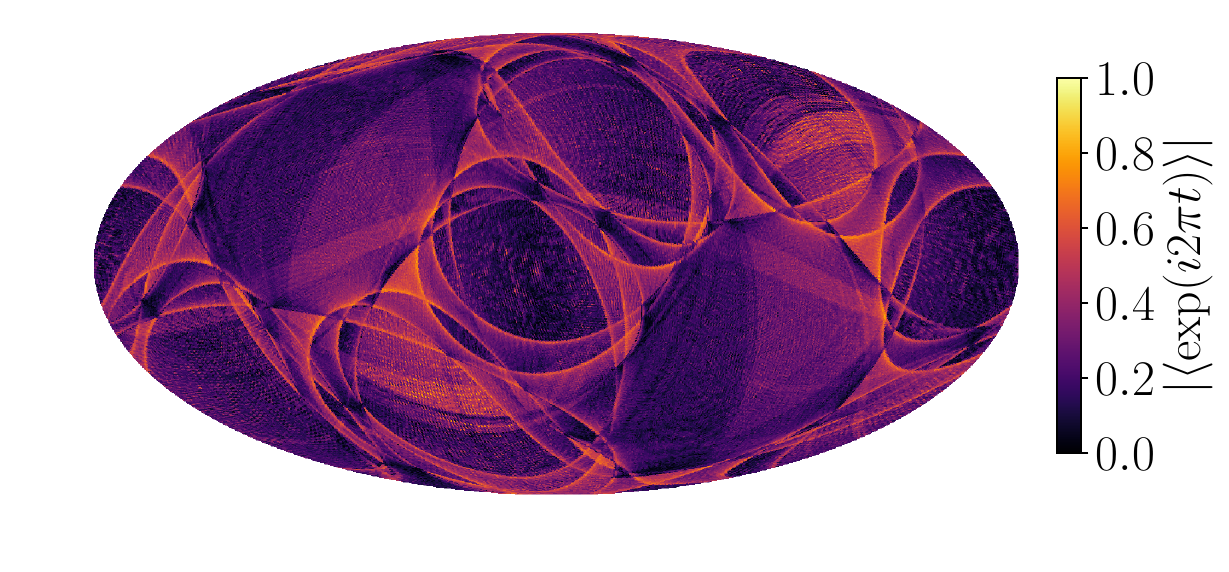} 
    \vspace*{-5mm}
    \caption{Measuring parallax requires position measurements throughout the year. Scans clustered at one time of year (light yellow) will produce a weaker parallax measurement than a spread of scans through the year (dark purple).} 
    \label{fig:slstats5}
    \end{subfigure}
     \caption{The precision with which \gaia measures 5D astrometry is heavily dependent on the number of FoV transits, scan times and directions, obtained from the scanning law. The plots provide some central summary statistics of the scanning law as a function of position on the sky in Galactic coordinates on HEALPix level 7.}
     \label{fig:slstats}
\end{figure}

\subsection{The Scanning Law}

The \gaia spacecraft is in orbit around the Lagrange point 2 (L2), orbiting the Sun in phase with the Earth. The spacecraft spins with a 6 hour period around a central axis which precesses with an aspect angle of 45 deg around the pointing connecting the satellite and the Sun, with a 63 day period. This is similar to a spinning top which has been left long enough to wobble. The orbit of the spacecraft around the sun adds a third axis of rotation.

Perpendicular to the spin axis, two fields of view (FoV) observe in directions separated by 106.5 deg. The direction in which each FoV is pointing at any point in time throughout \gaia's observing period is the scanning law.

The \gaia DR2 observing period runs from July 25 2014 (10:30 UTC) until May 23 2016 (11:35 UTC). The scanning law for DR2 is published by DPAC and refined by \citet{Paper1} and \citet[hereafter \citetalias{Paper3}]{Paper3}. 
Whilst this tells us where \gaia was pointing, it doesn't tell us whether Gaia was obtaining useful scientific measurements that contributed to the published data products. Many time periods in the DR2 window did not result in measurements which contributed to the \gaia astrometry as discussed in \citetalias{Paper3}. 

In this paper we only include the scanning law in the OBMT\footnote{Onboard Mission Time (OBMT) is the timing system used in \gaia and is normalised such that OBMT is 0 in October 2013 and increments by 1 for every revolution of the \gaia satellite which corresponds to 6 hours}  interval 1192.13–3750.56 rev  \citep[hereafter \citetalias{Lindegren2018}]{Lindegren2018} which removes the Ecliptic Polar Scanning Law, an initial calibration phase of \gaia which contributed to the published photometry but not astrometry. DPAC have published a series of additional gaps in astrometry data taking\footnote{\url{https://www.cosmos.esa.int/web/gaia/dr2-data-gaps}}. We remove any time spans of the scanning law for which the gap is flagged as `persistent'. Using the published Epoch Photometry for 550,737 variable sources \citep{Riello2018, Evans2018, Holl2018}, \citetalias{Paper3} constrained additional gaps which are persistent across all data products of \gaia DR2 which we also remove from the scanning law. Finally, \citetalias{Paper3} determines the probability of an observation being recorded and used in \gaia DR2 in 19 magnitude bins. These observation probabilities will be used to weight observations in the ASF in Section~\ref{sec:weights}.

\subsection{Taking Observations}

Both FoVs project source images onto a single panel of CCDs called the focal plane. On the focal plane there are 9 columns and 7 rows of CCDs, referred to as the astrometric field (AF), which measure the position of a source although the middle row only has 8 CCDs (because one of the 9 CCD positions is taken by a wave front sensor). As the spacecraft spins, stars track across the CCD panel in the along-scan direction and are observed with up to 9 astrometric CCDs during a single FoV transit  \citep[see e.g. Fig.~1 of][]{Lindegren2016}. Individual CCD measurements will be referred to as observations whilst a full track across the CCD panel is a scan (also referred to as a FoV transit). Before the AF, sources pass over the `Sky Mapper' (SM) CCD which triggers the initial detection and needs to be confirmed by the first AF CCD in order for any observations within the scan to successfully provide a measurement. Each observation records the position and apparent brightness of the source. If the source is recorded with $G<13$ by the SM, a 2D observation window is assigned measuring position in the along-scan (AL) direction and orthogonal across-scan (AC) direction. 
For fainter stars with $G>13$, only the AL position is recorded.

Observations are saved on-board \gaia in `Star Packets' grouped by apparent magnitude in 19 bins\citep[Table 1.10, Section 1.3.3][]{DocDR2ch1_2018}. The majority of data is uploaded to Earth, however some can be lost or deleted \citep[see Section 3.3][]{Prusti2016} changing the scanning law sampling for stars in different Star Packet magnitude bins.

After a first process of CCD signal level, background, and PSF/LSF calibration,  the data is input to the AGIS pipeline \citep{Lindegren2012} which uses an iterative linear regression algorithm to simultaneously fit the attitude of the spacecraft, a large number of calibration parameters, and the position, proper motion and parallax of all sources in \gaia DR2. We here provide a general description of how position, proper motion and parallax can be understood to depend on the nature of the observations a source receives, though in \gaia they are simultaneously solved from the offsets of all source observations with respect to its (iteratively improved) internal reference system.

The precision with which position, proper motion and parallax of a source can be measured is heavily dependent on magnitude (beyond $G>13$ the uncertainties will monotonically increase with magnitude), the number of observations taken, the scan directions of these observations and their distribution in time. More observations will produce a greater precision therefore sources in regions of the sky with the most scans as shown in Fig.~\ref{fig:slstats1} will have the best constrained astrometry. Notably, the Galactic center in the middle of the plot has received only $\sim 10$scans whilst the best observed regions of the sky are scanned over 100 times.

For the vast majority of sources, \gaia only measures position in the AL direction and even for 2D observations, the AL position constraint is much tighter than the AC measurement \citep{Lindegren2012}. Therefore North-South scans in equatorial coordinates will constrain declination, $\delta$ whilst East-West scans constrain right ascension, $\alpha$. Fig.~\ref{fig:slstats2} gives the mean direction of \gaia DR2 scans modulo $\pi$ such that a North-South and South-North scan appear the same with $\langle \phi\rangle = 0$. The mean direction is estimated from the argument of the mean vector
\begin{equation}
    \langle \phi \rangle = \frac{1}{2}\mathrm{arg}\left[\langle\exp(2i\phi)\right\rangle].
\end{equation}
This statistic is published for sources in \gaia EDR3 as \textsc{scan\_direction\_mean\_k2}. Darker areas will have stronger $\delta$ constraints whilst lighter areas constrain $\alpha$ more tightly. The difference in accuracy in right ascension and declination depends on the clustering of scan directions. The absolute value of the mean scan vector, $\left|\langle\exp(2i\phi)\rangle\right|$ which will be $\sim1$ for heavily clustered scans and $\sim0$ for a spread of scan directions. This is shown in Fig.~\ref{fig:slstats3} where light areas will strongly constrain position in the mean scan direction but only provide a weak constraint in the perpendicular direction whilst dark regions have a spread of scan directions and therefore won't show a strong direction preference. This statistic is also published in \gaia EDR3 as \textsc{scan\_direction\_strength\_k2}.

Constraints on the source proper motion come from measuring the position of a source at multiple different epochs and estimating the rate of change. A larger spread of observation times will produce a tighter proper motion constraint. Fig.~\ref{fig:slstats4} shows the standard deviation of observation times with light regions producing tighter proper motion constraints whilst dark regions produce a weaker constraint.

Finally, source parallax is estimated from the apparent motion of a source relative to the background of distant sources due to \gaia's motion around the sun on a one year period. A larger spread of observations throughout the year will produce a tighter constraint on the source parallax. The position of an observation in the yearly solar orbit is described by the complex vector $\exp(2\pi i t)$. As with the scan direction, the clustering of observations in the year is estimated from the absolute value of the mean vector $\left|\langle\exp(2\pi i t)\rangle\right|$. If observations are heavily clustered at one time of year the absolute mean will be close to 1, shown by lighter areas of Fig.~\ref{fig:slstats5}, and only a weak constraint on parallax will be achieved. Values close to 0 have well spread observations throughout the year and therefore provide a stronger constraint on parallax.


\subsection{Data}

The previous sections have provided a qualitative prediction of \gaia's expected performance as a function of position on the sky. In the following sections we'll produce a quantitative estimate of the predicted precision with which \gaia can measure source astrometry as a function of position on the sky and apparent magnitude. 

\gaia DR2 \citep{Prusti2016, Brown2018} provides 5D astrometry for 1,331,909,727 of the 1,692,919,135 source in the full DR2 catalogue. To test our predictions, we'll use the full \gaia DR2 source catalogue and 5D astrometry sample.

\section{Method}
\label{sec:method}

As input AGIS takes the 1D measurement of the position of each source in the AL and, for bright sources, also AC direction. For bright sources, \gaia produces a 2D observation however AGIS assumes the constraints in the AL and AC directions are uncorrelated treating them as independent 1D observations. 
We make the same assumption in this work. This is a gross simplification of all the steps which AGIS takes -- for instance, calibrating the satellite attitude noise -- however it allows for a very appealing and tractable derivation of the ASF from the available data.

We will proceed with four key assumptions:
\begin{itemize}
    \item The 1D position measurement uncertainty is Gaussian.
    \item Individual measurements, including AL and AC measurements from the same observation, are independent and uncorrelated. As the AGIS pipeline uses the same assumption, this will not produce any discrepancy between our predictions and the published \gaia astrometry.
    \item The position measurement uncertainty is a function of source apparent magnitude at the time of observation only. Any dependence of the observation precision of the satellite as a function of time for a given apparent magnitude is neglected which we justify in Appendix~\ref{sec:timedep}. This also assumes that the measurement uncertainty is colour independent.
    \item Astrometric parameters of different sources are assumed to be independent. In reality measurements of different sources can be considered independent, however due to the joint estimation of the attitude and geometric calibration from the same set of observations, the posterior astrometric parameters will be correlated. Pre-launch estimates by \citet{Holl2010} predicted correlations of only a fraction of a percent for sources separated by less than one degree (in a fully calibrated AGIS solution dominated by photon noise). DR1 \citep[see Sec. D.3 of][]{Lindegren2016} seems to be well above that with correlation as high as perhaps 0.25 at separations up to $\sim1$~deg, though much smaller on longer scales. Studies on the quasar sample in DR2 \citep[see Sect.~5.4 of][]{Lindegren2018} show still very large covariances as small spatial scales ($<0.125$~deg) and milder effects over larger spatial scales. With each successive data release it is expected that these spatial correlations will shrink, though they will never be zero, especially at small scales.
\end{itemize}
Throughout this paper we also only consider sources with constant magnitude to keep the results simple and tractable however the formalism is easily generalisable to variable sources. Over this section we derive the ASF of \gaia DR2. As many different variables are introduced, we refer the reader to Table~\ref{tab:notation} to clarify our notation.

\subsection{Astrometry from linear regression}
\label{sec:ast5d}

\begin{table}
\begin{tabular}{|| l | l | c || }
\hline
\hline
$\mathbf{x}$ & $\mathbb{R}^N$ & True AL position of source at observation time.\\
$\mathbf{x}'$ & $\mathbb{R}^N$ & Measured AL position of source at observation time.\\
$\mathbf{r}$ & $\mathbb{R}^5$ & Source astrometry.\\
$M$ & $\mathbb{R}^{N\times5}$ & Design matrix of astrometric solution.\\
$K$ & $\mathbb{R}^{N\times N}$ & Expected measurement covariance.\\
$C$ & $\mathbb{R}^{5\times5}$ & Published astrometry covariance.\\
$\Sigma$ & $\mathbb{R}^{5\times5}$ & Expected astrometry covariance ($\Sigma=\rho\,\Phi$).\\
$\rho$ & $\mathbb{R}$ & Magnitude dependence of ASF.\\
$\Phi$ & $\mathbb{R}^{5\times5}$ & Spatial dependence of ASF from scanning law.\\
\hline
\hline
\end{tabular}
\caption{Notation followed in linear regression.}
\label{tab:notation}
\end{table}

\gaia's goal for each source in the astrometry catalogue is to measure the five parameter astrometric solution, $\mathbf{r}\in \mathbb{R}^5$, consisting of the positions, proper motions and parallax. The AGIS pipeline estimates the astrometry of sources through linear regression on all observations of a single source in a step called source update \citep[see Sect.~5.1 of ][]{Lindegren2012}. We will use the same technique to determine the expected precision for a simple point source as a function of apparent magnitude and position on the sky.

Take $N$ observations of a source at times $t_i$ with the scan direction $\phi_i$ where $i\in\{1,\dots,N\}$. The on-sky positions of the source at times $t_i$ in ICRS coordinates are $(\alpha_i,\delta_i)$. The source position relative to the solar system barycenter at the reference epoch \citepalias[J2015.5 for \gaia DR2 ][]{Lindegren2018} is $\alpha_0,\delta_0$. 
The position at time $t_i$ will be a linear combination of the position at a reference epoch with the proper motion and parallax motion. The offset due to parallax motion is given by
 \begin{align}
    &\Delta\alpha_i\cos\delta_i = -\varpi\left(-X_i\sin\alpha_0 + Y_i\cos\alpha_0\right) = \varpi \Pi_{\alpha_i} \\
    &\Delta\delta_i = -\varpi\left(-X_i\cos\alpha_0\sin\delta_0 - Y_i\sin\alpha_0\sin\delta_0 + Z_i\cos\delta_0\right) = \varpi \Pi_{\delta_i} 
\end{align}
where $X_i,Y_i,Z_i$ are the barycentric coordinates of \gaia at time $t_i$ and $\varpi$ is the parallax of the source. We have assumed the parallax and proper motion are small enough such that the parallax ellipse is only dependent on the source reference epoch position which keeps the system of equations linear.

Therefore the position of the source at time $t_i$ will be given by
\begin{align}
    &\alpha_i^* = \alpha_0^* + \mu_{\alpha^*} t_i + \varpi\Pi_{\alpha_i} \\
     &   \delta_i = \delta_0 + \mu_\delta t_i + \varpi\Pi_{\delta_i}
\end{align}
where $\mu_\alpha, \mu_\delta$ is the source proper motion, $t_i$ is the time relative to the reference epoch and we use the notation $\alpha^* = \alpha\cos(\delta)$ and $\mu_{\alpha^*} = \mu_\alpha \cos(\delta)$.

Writing this set of linear equations out in matrix notation:
\begin{align}
    \begin{bmatrix}
           \alpha_1^* \\
           \delta_1 \\
           \vdots \\
           \alpha_N^* \\
           \delta_N
         \end{bmatrix}
         = 
           \begin{bmatrix}
         1&0&\Pi_{\alpha_1}&t_1&0  \\
         0&1&\Pi_{\delta_1}&0&t_1  \\
          && \vdots&& \\
         1&0&\Pi_{\alpha_N}&t_N&0  \\
         0&1&\Pi_{\delta_N}&0&t_N  \\
         \end{bmatrix}
                   \begin{bmatrix}
         \alpha_0^*\\
         \delta_0\\
         \varpi\\
         \mu_{\alpha^*}\\
         \mu_\delta
         \end{bmatrix}.  
         \label{eq:equatmatrix}
\end{align}

Our measurables are 1D positions in either the AL or AC direction of the \gaia focal plane. This is given by ${x_i = \alpha_i^* \sin\phi_i + \delta_i \cos\phi_i}$ where the scan position angle, $\phi_i$ is the scan direction of \gaia at the observation time for AL observations (shifted by $\pi/2$ for AC observations) and is defined such that $\phi=0^{\circ}$ in the direction of local Equatorial North, and $\phi=90^{\circ}$ towards local East\footnote{\url{https://www.cosmos.esa.int/web/gaia/scanning-law-pointings}}.

Substituting $x_i$ into Eq.~\ref{eq:equatmatrix}:
\begin{align}
    \begin{bmatrix}
           x_1 \\
           x_2 \\
           \vdots \\
           x_N 
         \end{bmatrix}
        & = 
           \begin{bmatrix}
         s_1&c_1&\Pi_{\alpha_1}s_1+\Pi_{\delta_1}c_1&t_1s_1&t_1 c_1 \\
         s_2&c_2&\Pi_{\alpha_2}s_2+\Pi_{\delta_2}c_2&t_2s_2&t_2 c_2 \\
          && \vdots&& \\
         s_N&c_N&\Pi_{\alpha_N}s_N+\Pi_{\delta_N}c_N&t_Ns_N&t_N c_N \\
         \end{bmatrix}
                   \begin{bmatrix}
         \alpha_0^*\\
         \delta_0\\
         \varpi\\
         \mu_{\alpha^*}\\
         \mu_\delta
         \end{bmatrix}  \nonumber\\
        & = \mathbfss{M} \mathbf{r}
\end{align}
where $c_i = \cos\phi_i$ and $s_i = \sin\phi_i$, $\mathbfss{M} \in \mathbb{R}^{N\times5}$ is the design matrix for the linear equations and $\mathbf{r} \in \mathbb{R}^{5}$ is the vector of astrometric parameters.

Assuming Gaussian measurement uncertainty for both AL and AC measurements and assuming all observations are independent, the observed source positions are distributed $\mathbf{x}'\sim\mathcal{N}\left(\mathbf{x}, \mathbfss{K}\right)$ where the covariance matrix $\mathbfss{K} = \mathrm{diag}\left[\sigma_1^2, \dots, \sigma_N^2 \right]$. This measurement covariance implicitly assumes all observations are independent and uncorrelated, one of our key assumptions also adopted in AGIS.

Following standard linear least squares regression \citep{Hogg2010}, the astrometric uncertainty covariance matrix of the inferred $\mathbf{r}$ is given by
\begin{align}
    \Sigma^{-1} = \mathbfss{M}^\mathrm{T} \mathbfss{K}^{-1} \mathbfss{M}.
\end{align}
Expanding this out in terms of all scan angles, the full inverse covariance matrix is given by
\begin{align}
\label{eq:precision}
    \Sigma^{-1} = \sum_{i=1}^{N} \frac{1}{\sigma_i^2} \mathbfss{A}_i 
\end{align}
with
\begin{align}
\label{eq:Amatrix}
    \mathbfss{A}_i = \begin{bmatrix}
s_i^2 &  s_ic_i& s_i\Pi_i& s_i^2t_i& s_ic_it_i \\
s_ic_i& c_i^2& c_i\Pi_i & c_is_it_i& c_i^2t_i\\
s_i\Pi_i& c_i\Pi_i& \Pi_i^2& s_it_i\Pi_i& c_it_i\Pi_i \\
s_i^2 t_i& s_ic_it_i& s_it_i\Pi_i& s_i^2t_i^2& s_ic_it_i^2 \\
s_ic_i t_i& c_i^2t_i& c_it_i\Pi_i& s_ic_it_i^2& c_i^2t_i^2 \\
         \end{bmatrix} \nonumber \\
\end{align}
where $\Pi_i = \Pi_{\alpha_i}s_i+\Pi_{\delta_i}c_i$.

Eq.~\ref{eq:precision} assumes that every scan of a source will produce a detection which contributes to the astrometric solution. Even after removing gaps in the scanning law, there are periods of time and magnitudes which are less likely to result in good astrometric observations. We need to account for the efficiency of \gaia observations.

\subsection{Scan Weights}
\label{sec:weights}

As \gaia scans a source, up to 9 observations are taken with the 9 astrometric-field CCD columns. There are two ways in which observations may not be propagated to the astrometric solution. If a source is not detected and confirmed by the SM and first AF CCDs and allocated a window, none of the CCDs in the scan will produce a successful detection. Secondly, an individual CCD observation may either not be taken or the measurement may be down-weighted in the astrometric solution. There are many reasons why this might happen such as stray background light, attitude calibration or the source simply passing through the small gaps between CCD rows. Accounting for these processes, the astrometric precision matrix may be approximated as
\begin{equation}
    \Sigma^{-1} = \sum_{i=1}^N \frac{y_i}{\sigma_i^2} \mathbfss{A}_i
\end{equation}
where $y_i\sim\mathrm{Bernoulli}(\xi_i) \times \mathrm{Binomial}(9, \theta_i)$ where $\xi_i$ is the fraction of scans used in the astrometric solution and $\theta_i$ is the probability of a CCD producing a successful observation. The binomial distribution assumes that a CCD observation is either successful or not therefore only allowing a full weight or zero weight. The weight formula in the AGIS pipeline \citep[Eq. 66, ][]{Lindegren2012} does allow for non-discrete weights however we anticipate that this will have a small effect on our results.
Assuming that the event of a successful scan or observation are independent events, the expected value of the weights is given by
\begin{align}
    w_i = \mathbb{E}\left[y_i\right] &= \mathbb{E}\left[\mathrm{Bernoulli}(\xi_i)\right]\times\mathbb{E}\left[\mathrm{Binomial}(9, \theta_i)\right]\nonumber\\
    &= \xi_i \times 9\theta_i.
\end{align}
Therefore the expected astrometric precision is given by
\begin{align}
\label{eq:w_precision}
    \mathbb{E}\left[ \Sigma^{-1}\right] = \sum_{i=1}^{N} \frac{w_i}{\sigma_i^2} \mathbfss{A}_i.
\end{align}

For sources with $G>13$, \gaia only measures a 1D position in the AL direction however for bright sources, $G<13$ and AC measurement is also taken. Following the method in \citet{Lindegren2012}, we treat the AL and AC observations as independent 1D measurements such that Eq.~\ref{eq:w_precision} expands out to
\begin{align}
\label{eq:wacprec}
    \mathbb{E}\left[\Sigma^{-1}\right] = &  \sum_{i=1}^{N_{\mathrm{AL}}} \frac{w_i^\mathrm{AL}}{\sigma_{\mathrm{AL},i}^2} \mathbfss{A}_i+   \sum_{i=1}^{N_{\mathrm{AC}}} \frac{w_i^\mathrm{AC}}{\sigma_{\mathrm{AC},i}^2} \mathbfss{A}_i.
\end{align}

\begin{figure}
  \centering
  \includegraphics[width=0.495\textwidth]{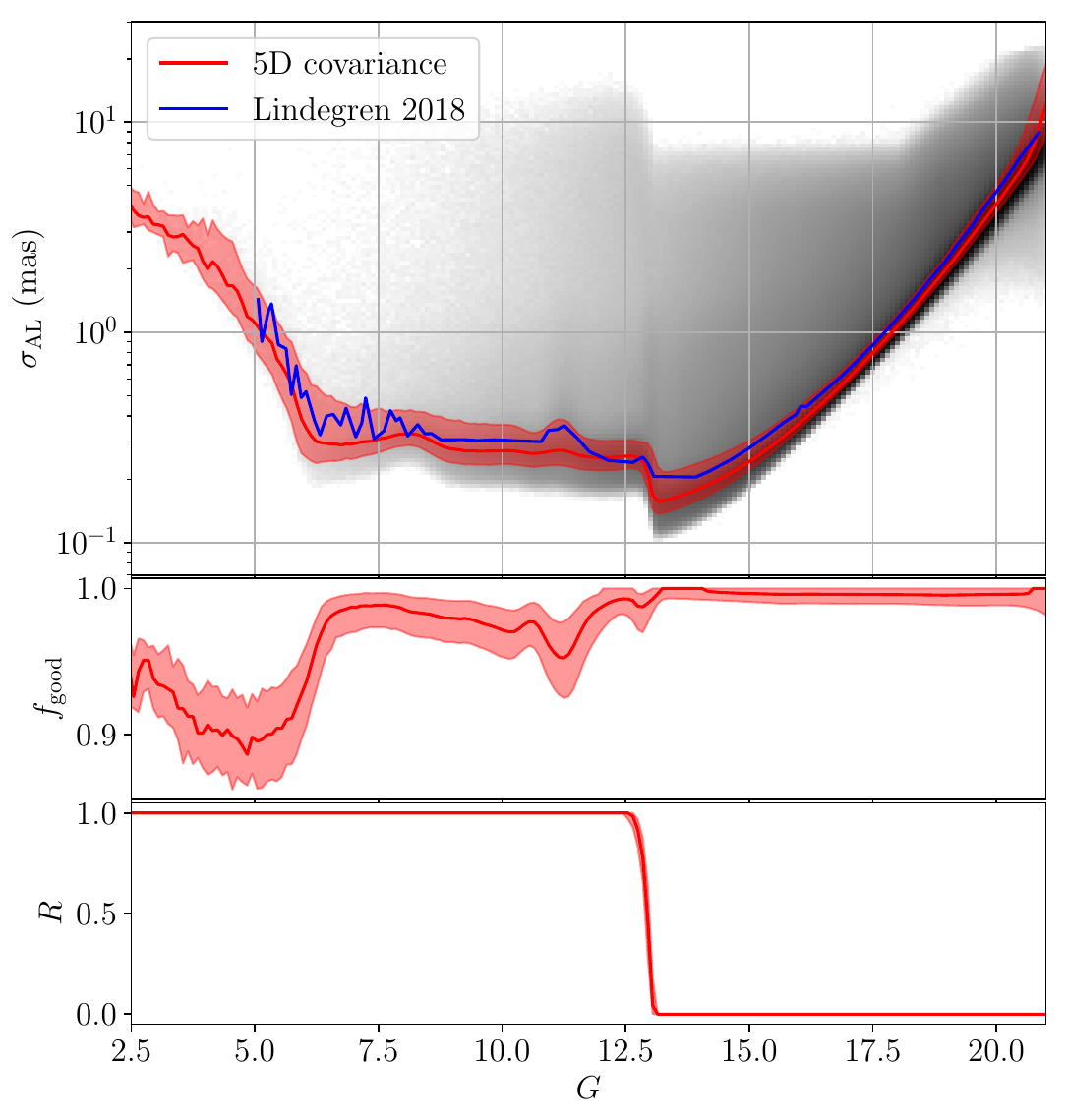}
  \caption[]{The magnitude dependence of the ASF is a function of the AL astrometric uncertainty $\sigma_\mathrm{AL}$ (top), the fraction of photometric observations which generate good astrometric observation used in the AGIS pipeline $f_\mathrm{good}$ (middle) and the ratio of AC to AL observations $R$ (bottom). In all cases the median and $16^\mathrm{th}$ to $84^\mathrm{th}$ percentiles of the \gaia DR2 astrometry sample are given by the red sold line and shaded area respectively. The distribution of $\sigma_\mathrm{AL}$ (black histograms, log normalised) extends high above the median due to source excess noise.}
   \label{fig:sigma_al}
\end{figure}

In \citet{Paper3} the fraction of scans, $f(t_i)$ which contribute to the \gaia photometry is estimated in Star Packet magnitude bins as a function of time in DR2. Due to their separate pipelines, the probability of an observation contributing to the astrometric solution will differ from the photometry. To determine the astrometry weights, we renormalise the photometry scan fraction using the published number of astrometric detections used, \textsc{astrometric\_n\_good\_obs\_al}
\begin{align}
    w_i^\mathrm{AL} &=9\times\frac{62}{63}\,f(t_i)  \left\langle\frac{\textsc{astrometric\_n\_good\_obs\_al}} {9\times\frac{62}{63}\sum_{i=1}^{N_\mathrm{scan}}f(t_i)} \right\rangle_G \nonumber \\
    &= \upsilon \, f(t_i)\,  f_\mathrm{good}(G).
\end{align}
where $w_i^\mathrm{AL}$ is the weight for AL source observations since we have renormalised by the number of good AL observations used in the astrometry. The multiplication of the scan fraction by $\upsilon = 9\times\frac{62}{63}$ converts the scan fraction to average number of observations. There are 9 columns and 7 rows of CCDs in the astrometric field of \gaia however one CCD is replaced by a wave front sensor hence only 62 are left. $f_\mathrm{good}(G)$, shown in the middle panel of Fig.~\ref{fig:sigma_al}, is above 90\% across the most magnitudes and only significantly deviates from 100\% at the bright end. 

For a given source, the number of AL and AC observations is published in \gaia DR2 as $\textsc{astrometric\_n\_obs\_al}$ and $\textsc{astrometric\_n\_obs\_ac}$. These statistics do not account for down-weighting of observations in the astrometry pipeline, however, assuming the AL and AC measurements of the same observations are equally likely to be down-weighted, the ratio between the numbers will be unaffected. $R=\frac{\textsc{astrometric\_n\_obs\_ac}}{\textsc{astrometric\_n\_obs\_al}}$ gives the fraction of observations which produce an AC measurement. The bottom panel of Fig.~\ref{fig:sigma_al} shows that observations with $G<13$ produce AC measurements whilst $G>13$ do not. We use this fraction to relate the observation weights ${{w_i^\mathrm{AC} = R(G) w_i^\mathrm{AL}}}$. In truth, the scan fraction, $R$, may be a weak function of position on the sky at bright magnitudes due to crowding causing problems with window assignment. However, as we'll see in the following section, the contribution from AC observations to the astrometric precision is $\sim 3\%$ compared to the AL contribution and so any weak uncertainty in $R$ will have a small impact on the estimated precision.

\begin{figure}
  \centering
  \includegraphics[width=0.495\textwidth]{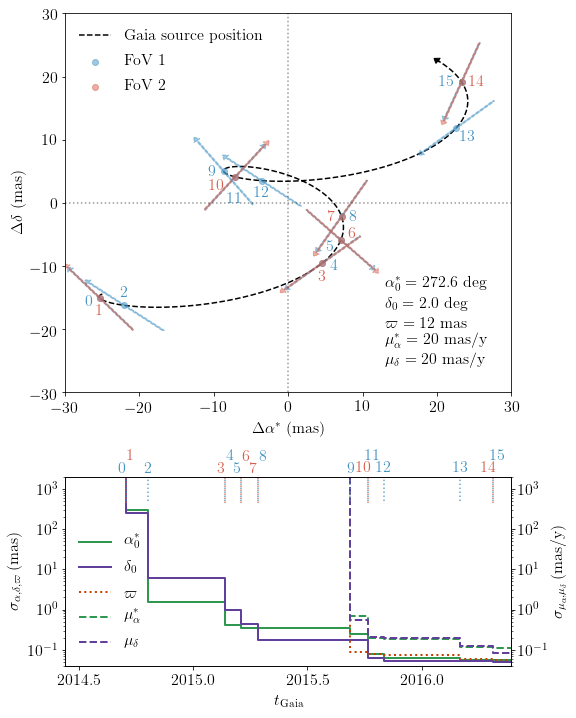}
  \caption[]{\textbf{Top}: Observed from L2, any source on the sky follows a curved track given by the combination of source proper motion and apparent parallax ellipse as represented here by the black-dashed line for a source with $\mu_\alpha^*=20\,\mathrm{mas}/\mathrm{y}$, $\mu_\delta=20\,\mathrm{mas}/\mathrm{y}$ and $\varpi=10\,\mathrm{mas}$ located near the galactic bulge with $l=30$ deg, $b=10$ deg. \gaia scans this position on the sky 15 times in the DR2 time frame shown by blue and red arrows for FoV1 and FoV2 respectively. \textbf{Bottom}: Each scan, marked by the vertical red and blue dashed lines, contributes to the 5D astrometry constraints. The expected uncertainty on each astrometry parameter is shown for a source with $G=16$ and therefore $\sigma_\mathrm{AL}=0.37$ mas and reduce with each subsequent scan. When fewer than 6 visibility periods are observed, only $\alpha_0^*$ (green solid) and $\delta_0$ (purple solid) are shown with priors placed on all parameters. With at least 6 visibility periods, uncertainties are also given for $\mu_\alpha^*$ (green dashed), $\mu_\delta$ (purple dashed) and $\varpi$ (red dotted). 
  }
   \label{fig:example_scans}
\end{figure}
\begin{figure*}
  \centering
  \includegraphics[width=\textwidth]{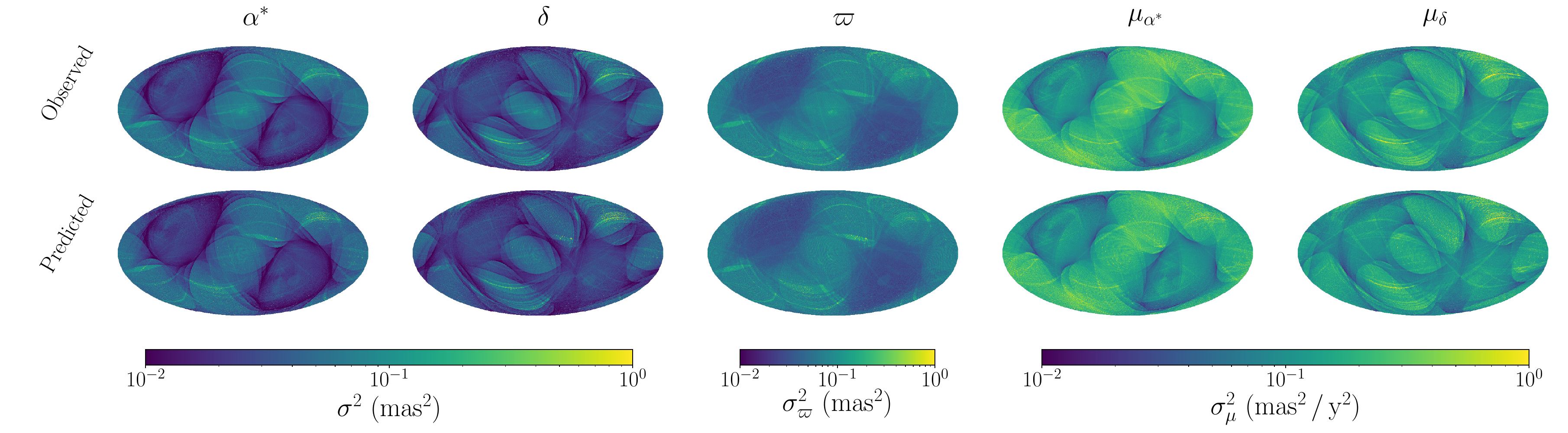}
  \caption[]{The predicted variance in each component of the astrometry for the $G\in[18.1,19.0]$ magnitude bin (bottom row) matches the median variances for observed \gaia astrometry sources in the same magnitude range (top row) in HEALPix level 7 bins shown in Galactic coordinates. In all plots, the lighter shades near the ecliptic plane show increased variance due to lack of scans in \gaia DR2. The top row is similar to \citetalias[][Fig. B3]{Lindegren2018} where instead of showing individual components, they instead show the position and proper motion semi-major axes.} 
   \label{fig:5dprec_diag}
\end{figure*}
\begin{figure*}
  \centering
  \includegraphics[width=\textwidth]{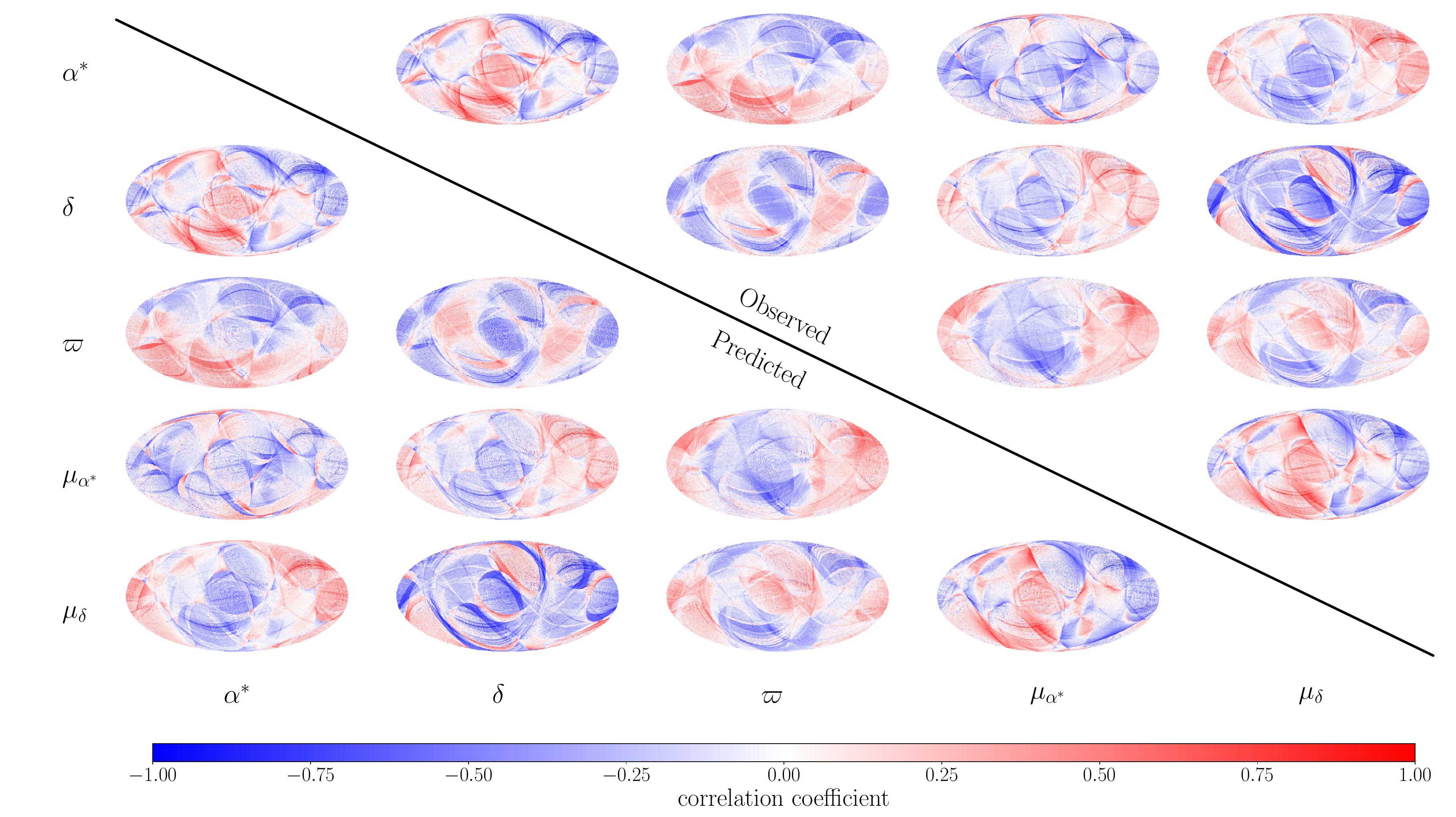}
  \caption[]{The predicted correlation coefficients for $G\in[18.1,19.0]$ (lower triangle) match up well to the correlation coefficients of the median covariances from the \gaia astrometry (upper triangle) in HEALPix level 7 bins shown in Galactic coordinates. The detailed and complex structure in the correlation coefficients is driven by the directions and time separations between subsequent scans of the same position on the sky. The $\varpi-\mu_\alpha^*$, $\varpi-\mu_\delta$ and $\mu_\alpha^*-\mu_\delta$ in the upper triangle correspond to \citetalias[][Fig. B5]{Lindegren2018} although at a different magnitude.}
   \label{fig:5dprec_corr}
\end{figure*}
\begin{figure*}
  \centering
  \includegraphics[width=\textwidth]{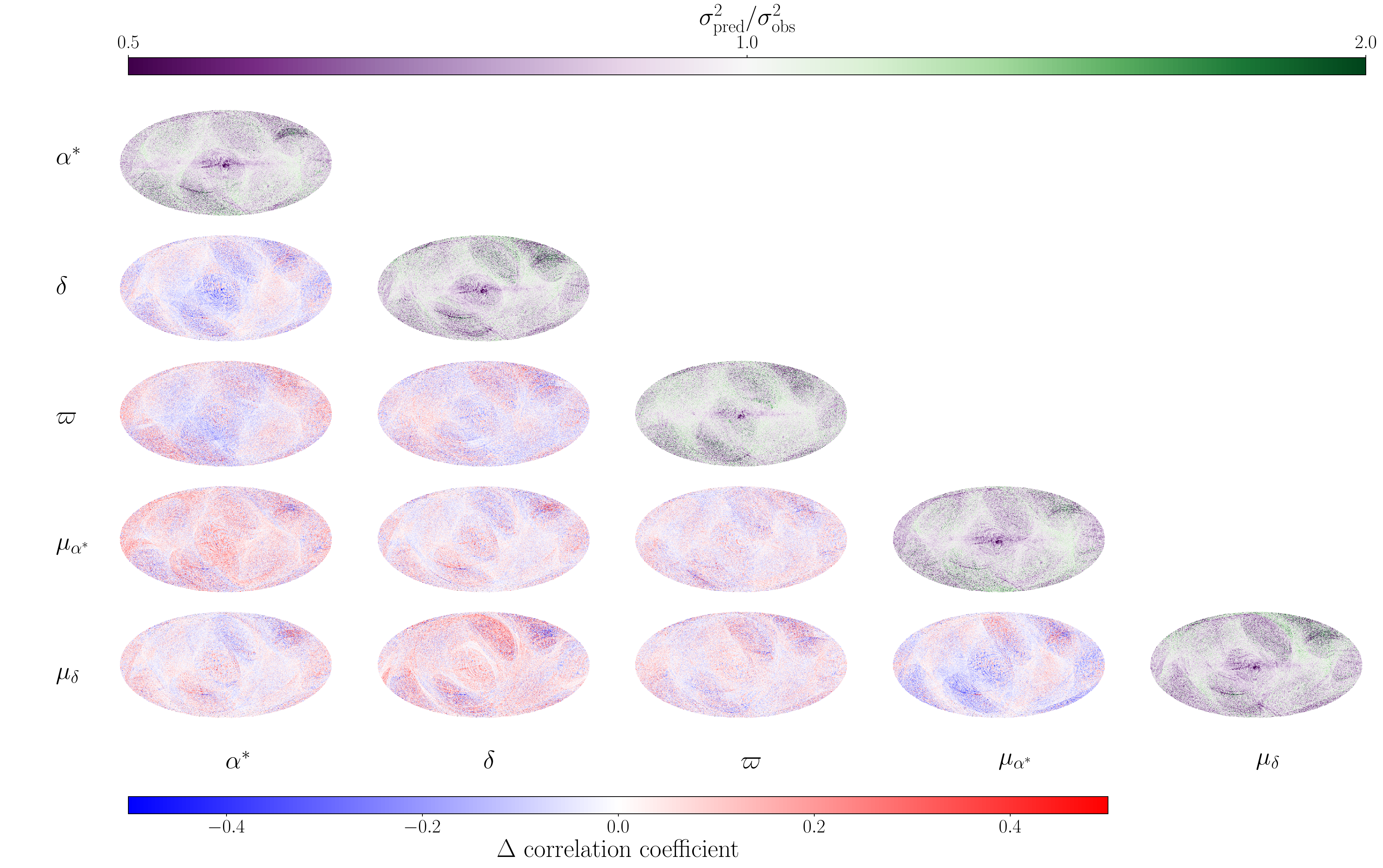}
  \caption[]{The ratio of predicted to median observed variance for $G\in[18.1,19.0]$ (diagonal)  in HEALPix level 7 bins shows structure in the bulge where the prediction has underestimated the observed variance driven by crowding of sources and lack of observations in \gaia DR2. The difference between predicted and median observed correlation coefficients (lower triangle) shows no strong structural bias in correlation. All projections are in Galactic coordinates.}
   \label{fig:5dprec_comp}
\end{figure*}
\begin{figure*}
  \centering
  \includegraphics[width=\textwidth]{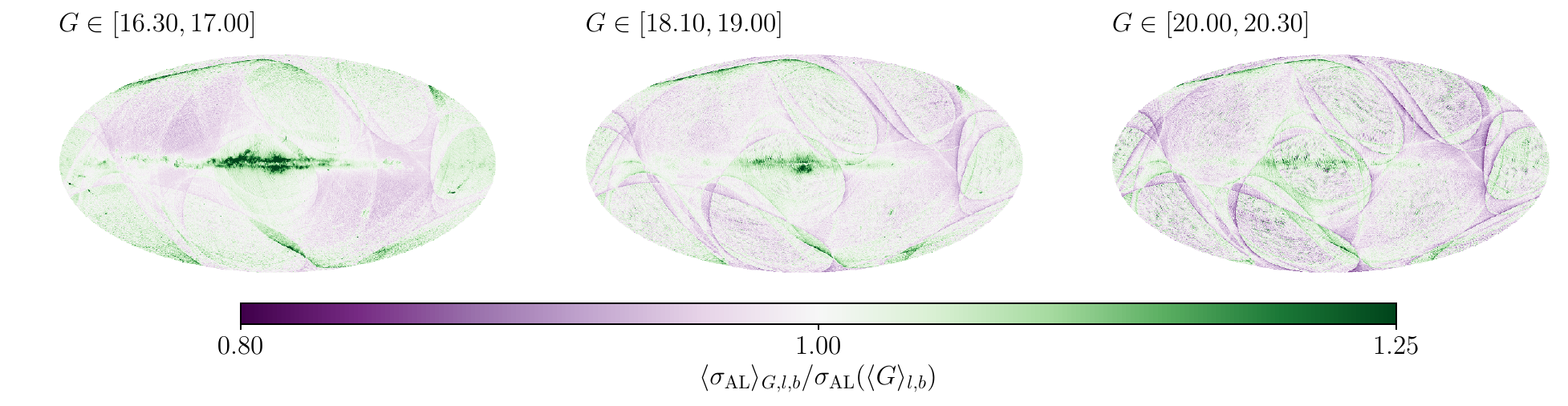}
  \caption[]{The ratio of the median $\sigma_\mathrm{AL}$ in HEALPix level 7 bins to the value of $\sigma_\mathrm{AL}$ evaluated at the median magnitude of stars in the HEALPix bins highlights any dependence of $\sigma_\mathrm{AL}$ on sky position in Galactic coordinates. Particularly in the highest source density regions of the Galactic plane and bulge at brighter magnitudes, sources have significantly higher astrometric measurement uncertainty than the average across the sky.}
   \label{fig:sigmaal_hpx}
\end{figure*}
\begin{figure}
  \centering
  \includegraphics[width=0.495\textwidth]{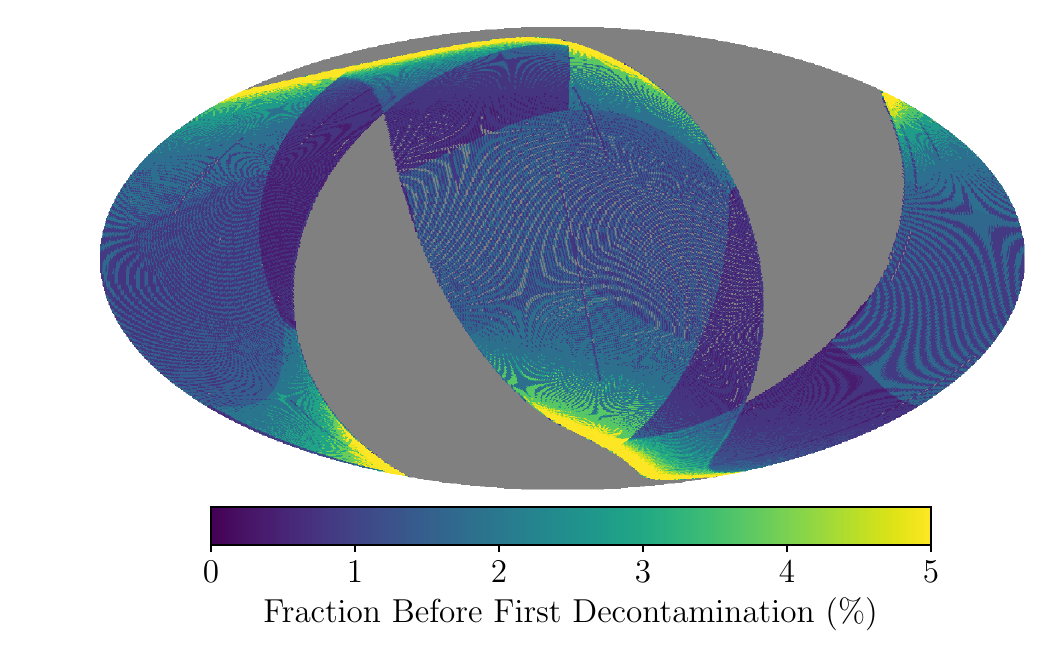}
  \caption[]{The fraction of scans of any position on the sky which occurred before the first decontamination event, shown here as a percentage in HEALPix 7 bins in Galactic coordinates, corresponds to regions of the sky with enhanced measurement uncertainty.}
   \label{fig:frac_decon1}
\end{figure}

\subsection{Centroid error}
\label{sec:centroid}

The centroid error in the AL and AC directions, $\sigma_i^\mathrm{AL}, \sigma_i^\mathrm{AC}$ is a function of the spacecraft instrumentation and apparent brightness of the source due to photon shot noise. For the remainder of this section, we assume that all CCDs in the astrometric field of the CCD panel have similar noise properties. We also assume that this performance is time independent and does not depend on the position of the source on the plane. Changes to the spacecraft such as mirror condensation and micrometeoroid impacts mean that the performance of the space craft is not perfectly time independent however we demonstrate in Appendix~\ref{sec:timedep} that the dependence is small compared to the scatter of individual measurements using the epoch photometry.

Therefore we assume that all AL observations of a single source have the same precision and likewise for all AC observations such that Eq.~\ref{eq:wacprec} becomes
\begin{align}
\label{eq:acprec}
    \mathbb{E}\left[\Sigma^{-1}\right] = & \frac{1}{\sigma_\mathrm{AL}(G)^2} \sum_{i=1}^{N_{\mathrm{AL}}} w_i^\mathrm{AL} \mathbfss{A}_i+  \frac{1}{\sigma_\mathrm{AC}(G)^2} \sum_{i=1}^{N_{\mathrm{AC}}} R(G)w_i^\mathrm{AL} \mathbfss{A}_i
\end{align}

\citet[Table 1]{Lindegren2012} gives the ratio of AC to AL error for bright sources as typically ${{\psi=520/92}}$ 
such that ${{\sigma_\mathrm{AC} = \psi\sigma_\mathrm{AL}}}$, although we note that this was only a pre-launch estimate and the true calibrated uncertainty is likely marginally different. Substituting into the expected precision
\begin{align}
\label{eq:al_prec}
    \mathbb{E}\left[\Sigma^{-1}\right] = & \frac{1}{\sigma_\mathrm{AL}(G)^2} \left(1 + \frac{R(G)}{\psi^2}\right) \sum_{i=1}^{N_{\mathrm{AL}}} w_i^\mathrm{AL} \mathbfss{A}_i.
\end{align}

The final unknown in Eq.~\ref{eq:final_prec} is $\sigma_\mathrm{AL}$, the astrometric centroid error of AL observations. $\sigma_\mathrm{AL}$ was estimated from the \gaia published astrometry by \citet{Belokurov2020} using the formula $0.53\sqrt{N}\sigma_\varpi$ where $N$ was the number of AL observations used for the source astrometry published as $\textsc{astrometric\_n\_good\_obs\_al}$. $0.53$ was used as this empirically matched the published distribution in Fig.9 of \citetalias{Lindegren2018}. However $\sqrt{N}\sigma_\varpi$ is a strong function of position on the sky depending on scan directions and spread of observations throughout the year. This means that the running median as a function of magnitude will be heavily affected by where the given stars lie on the sky. 

For this work, we find a more mathematically motivated route to the scan variance. By summing up the first two diagonal terms of the inverse covariance matrix from Eq.~\ref{eq:Amatrix}, the dependence on the scan angle $\phi_i$ disappears.
\begin{align}
\mathbb{E}\left[\Sigma^{-1}\right]_{\alpha,\alpha} + 
\mathbb{E}\left[\Sigma^{-1}\right]_{\delta,\delta} = &\frac{1}{\sigma_{AL}(G)^2} \left(1 + \frac{R(G)}{\psi^2}\right)   \sum_{i=1}^N w_i^\mathrm{AL} (s_i^2 + c_i^2) \nonumber\\
= &\frac{1}{\sigma_{AL}(G)^2} \left(1 + \frac{R(G)}{\psi^2}\right)  \sum_{i=1}^N w_i^\mathrm{AL}
\end{align}
Therefore the AL astrometric error can be determined independent of position on the sky by substituting $\Sigma$ for the published covariance $C$ and rearranging in terms of $\sigma_\mathrm{AL}$.
\begin{equation}
\label{eq:sigmaal}
\sigma_{AL}^2(G) =  \left(1 + \frac{R(G)}{\psi^2}\right) \left\langle\frac{\textsc{astrometric\_n\_good\_obs\_al}}{(\mathbfss{C}^{-1})_{\alpha\alpha}+(\mathbfss{C}^{-1})_{\delta\delta}}\right\rangle_G
\end{equation}
where $\sum_{i=1}^N w_i^\mathrm{AL} = \textsc{astrometric\_n\_good\_obs\_al}$. As we will discuss in more detail in Section~\ref{sec:sigma5dmax}, the selection of the \gaia 5D astrometry sample included a cut on \textsc{astrometric\_sigma5d\_max}. Sources with large astrometric uncertainty would not receive 5D astrometry and therefore, particularly at the dim end, $\sigma_\mathrm{AL}$ would be biased low. To mitigate this, we calculate $\sigma_\mathrm{AL}(G)$ using all stars in \gaia DR2 with at least 6 \textsc{visibility\_periods\_used}. For sources without 5D astrometry we use the inverse of the published 2D astrometry covariance matrix as a proxy. This is a rough approximation and therefore we suggest that our results are only trusted out to $G\lesssim 20.5$ at which point the cut on \textsc{astrometric\_sigma5d\_max} becomes significant (see Section~\ref{sec:sigma5dmax}).

The distribution of $\sigma_\mathrm{AL}$ is shown in the top panel of Fig.~\ref{fig:sigma_al} demonstrating a relatively flat behaviour for $G<13$ where 2D observations are taken and time windows are truncated to avoid saturation. For $G>18$ the variance grows with magnitude due to photon shot noise. The red line gives the median value in $0.1$mag and we linearly interpolate this as a function of magnitude to estimate $\sigma_\mathrm{AL}(G)$. For reference, the grey-scale histograms are the $\sigma_\mathrm{AL}$ for 5D astrometry sources where the truncation for $\sigma_\mathrm{AL}\sim 10~\mathrm{mas}$ is caused by the \textsc{astrometric\_sigma5d\_max} cut. The blue line in the top panel of Fig.~\ref{fig:sigma_al} is the blue line from \citetalias[Figure 9][]{Lindegren2018}. Across most of the magnitude range, our estimate is lower than \citetalias{Lindegren2018} by $\sim10\%$. This is expected because we're actually calculating slightly different statistics. \citetalias{Lindegren2018} used the residuals of all AL observations relative to the best fit astrometric solution. In calculating the source astrometry, observations are assigned weights as a function of their residuals which disfavoured observations with large residuals from being used in the astrometric solution. Therefore the value of $\sigma_\mathrm{AL}$ inferred by \citetalias{Lindegren2018} will be higher than ours which has implicitly ignored large outliers. As our task in this paper is to predict the published 5D astrometry uncertainties, our formula for $\sigma_\mathrm{AL}$ is the appropriate one to use.

Finally, we can substitute in $w_i^\mathrm{AL}$ from Section~\ref{sec:weights}.
\begin{align}
\label{eq:final_prec}
    \mathbb{E}\left[\Sigma^{-1}\right] = & \frac{1}{\sigma_\mathrm{AL}(G)^2} \left(1 + \frac{R(G)}{\psi^2}\right)  f_\mathrm{good}(G) \sum_{i=1}^{N_{\mathrm{AL}}} \upsilon f(t_i) A_i \nonumber\\
    = & \rho(G) \, \Phi(l,b)
\end{align}
where we have defined 
\begin{equation}
    \rho(G) \equiv \frac{1}{\sigma_\mathrm{AL}(G)^2} \left(1 + \frac{R(G)}{\psi^2}\right) f_\mathrm{good}(G)
    \label{eq:rhodef}
\end{equation} 
as the magnitude dependent normalisation and ${{\Phi(l,b) \equiv \sum_{i=1}^{N_{\mathrm{AL}}} \upsilon f(t_i) \mathbfss{A}_i}}$ as the scanning law dependent matrix. $\Phi$ has a weak magnitude dependence as the fractions $f(t_i)$ change between the magnitude bins in which \gaia downloads data however, within any download bin, it is independent of magnitude.

\subsection{Astrometry Spread Function}
\label{sec:asf}

In the previous sections we have derived the expected precision, $\mathbb{E}\left[\Sigma^{-1}\right]$ for simple point sources as observed by \gaia. The DR2 data has been used to estimate $\sigma_\mathrm{AL}(G), R(G)$ and $f_\mathrm{good}(G)$ as running medians as a function of magnitude. $\Phi(l,b)=\sum_{i=1}^{N_\mathrm{AL}}f(t_i) \mathbfss{A}_i$ is a function of the scanning law only and has no dependence on the \gaia astrometry data. For the remainder of this paper, we will simplify the notation taking ${\Sigma=\mathbb{E}\left[\Sigma^{-1}\right]^{-1}}$ as the expected 5D astrometry covariance for a simple point source in \gaia.

For a point source moving without acceleration with true astrometric coordinates $\mathbf{r}$ observed in \gaia DR2, the expected measured astrometric coordinates will be drawn from a multivariate normal distribution with covariance $\Sigma$,
\begin{equation}
    \mathbf{r}' \sim \mathcal{N} ( \mathbf{r}, \Sigma(G,l,b) ).
\end{equation}
This normal distribution is the Astrometry Spread Function where $G,l,b$ are the apparent magnitude and position of the source on the sky. 

To demonstrate how the astrometry is fit in practice, we show the expected observations and astrometric uncertainty for a hypothetical source at $l=30$ deg, $b=10$ deg with apparent magnitude $G=16$ in Fig.~\ref{fig:example_scans}. The source is given proper motion $\mu_\alpha^* =20\,\mathrm{mas}/\mathrm{y}$, $\mu_\delta =20\,\mathrm{mas}/\mathrm{y}$ which produces a trajectory from South East to North West. Adding the parallax ellipse for $\varpi=12$ mas generates a spiralling apparent position observed by \gaia throughout DR2 given by the black-dashed line in the top panel of Fig.~\ref{fig:example_scans}.

\gaia scans this region of the sky 15 times in DR2 given by the blue and red arrows for scans from FoV1 and FoV2 respectively. Each scan improves the constraint on each of the five astrometry parameters the uncertainties for which are given in the bottom panel of Fig.~\ref{fig:example_scans}. \gaia selects sources for the 5D astrometry catalogue which have at least 6 \textsc{visibility\_periods\_used} where a visibility period is a group of observations separated by less than four days. Where fewer than 6 visibility periods have been observed the AGIS pipeline places priors on the astrometry derived in \citet{Michalik2015} and only the 2D position constraints are published. We replicate this using the same priors and only providing uncertainties for the $\alpha_0^*$ (green solid) and $\delta_0$ (purple solid) parameters before the sixth visibility period (9th scan). 

After the sixth visibility period, the priors were dropped and the uncertainties on $\mu_\alpha^*$ (green dashed), $\mu_\delta$ (purple dashed) and $\varpi$ (red dotted) parameters are also shown. For simplicity, this demonstration assumes all observations were successful and equally likely to contribute to the astrometry however as discussed in Section~\ref{sec:weights}, this is not always the case and this is corrected for by weighting observations.

\section{Results}
\label{sec:results}

To test that our method is producing reasonable covariance matrices, we compare our predictions with the published 5D astrometry covariances. From the \gaia DR2 astrometry sample we determine the median published covariance on a level 7 HEALPix grid \citep{Gorski2005} in the magnitude range $G\in[18.1, 19.0]$ which represents a single Star Packet bin in which the scan fractions, $f$ are unchanged.

We estimate the predicted covariance using the formula in Eq.~\ref{eq:final_prec} where $G$ is taken as the median apparent magnitude of stars in the given magnitude bin and HEALPix pixel. The scan angles and times are inferred at the central coordinates of the HEALPix pixel. All figures are shown in Galactic coordinates.

The diagonal elements of both the median observed and predicted covariance matrices are shown in Fig.~\ref{fig:5dprec_diag} demonstrating excellent agreement down to degree scales in all components. In all coordinates the variance is significantly enhanced in regions which have been scanned less in DR2, most notably around the Galactic bulge. Thin streaks of boosted variance on the sky correspond to time periods in \gaia DR2 where data was not taken due to mirror decontamination or other disruptive processes.

In Fig.~\ref{fig:5dprec_corr} we compare the correlation coefficients evaluated by dividing the off-diagonal covariance elements by the square root of the products of their respective variances. Correlation coefficients are less dependent on the number of observations, which has largely been divided out, and more on the scan directions and time variance leading to a more complex and varied structure on the sky. Again, the observed correlation (upper right triangle) and predicted correlation (lower left) show excellent agreement down to small scale variations.

Fig.~\ref{fig:5dprec_comp} provides a more direct comparison between the predicted and observed covariances. Diagonal elements give the ratio of predicted to observed variance. Across the vast majority of the sky, there is strong agreement with noise dominating in underscanned regions. Two features stand out in the variance ratios where the model has not fully captured the system. A streak of scans in the South East and North West show underestimated uncertainties from the model. The scans in \gaia responsible for this are constrained and discussed in Section~\ref{sec:sigma5dmax}. Secondly, the Galactic bulge also shows a significant systematic underestimate against the observed variance. This is not unexpected as high source crowding can cause single windows to be allocated to multiple sources generating spurious centroid positions. The third panel of \citetalias[][Fig. B.4]{Lindegren2018} shows the same issue but manifested in the \textsc{astrometric\_excess\_noise} of the source fits. 

We demonstrate this issue in Fig.~\ref{fig:sigmaal_hpx} where we show the median $\sigma_\mathrm{AL}(G,l,b)$ evaluated using Eq.~\ref{eq:sigmaal} with the median taken in every $0.1$ mag magnitude bin and HEALPix level 7 pixel divided by $\sigma_\mathrm{AL}(G)$ evaluated at the median magnitude of stars in the HEALPix pixel. From Section~\ref{sec:centroid}, we expect $\sigma_\mathrm{AL}$ to be independent of position on the sky which is a key assumption in our model. Across the sky $\sigma_\mathrm{AL}$ shows only weak dependence on the scanning law at less than $\sim 10\%$. However, particularly for brighter magnitude bins, $\sigma_\mathrm{AL}$ is not uniform over the sky as expected and is significantly higher in regions of the disk and bulge with the highest source density. This issue is further exacerbated for the bulge as it happens to reside in a region of the sky which has been scanned very few times by \gaia whereas the LMC and SMC which have been scanned more heavily show no clear signal. In future \gaia data releases, the Galactic bulge will likely receive significantly more scans reducing this issue.

Fig.~\ref{fig:sigmaal_hpx} also shows residual scanning law structure which is likely caused by the $\sim20\%$ variation in the instrument precision discussed in Appendix~\ref{sec:timedep}. For example, the green strips in the North East and South where $\sigma_\mathrm{AL}$ is systematically higher correspond to areas which received many observations before the first decontamination when the satellite measurement precision was at its worst as shown in Fig.~\ref{fig:sigmaf_time}. Fig.~\ref{fig:frac_decon1} shows the percentage of observations which took place before the first decontamination event in DR2 for which the highest regions match exactly with regions of the sky in Fig.~\ref{fig:sigmaal_hpx} with enhanced $\sigma_\mathrm{AL}$. The diagonal elements of Fig.~\ref{fig:5dprec_comp} show that these features are comparable to the background noise level and so are not of significant concern.

The off-diagonal elements of Fig.~\ref{fig:5dprec_comp} show the difference between predicted and observed correlation coefficients. The structure of the scanning law can be seen in white as the regions which are most heavily scanned will have the lowest uncertainty. There is some marginal bias in the $\alpha^*$ and $\delta$ components but this is small compared with the overall signal seen in Fig.~\ref{fig:5dprec_corr}.

From these results, we demonstrate that the ASF is accurate across the majority of the sky across all magnitudes at the $10\%$ level. However for bright sources ($G\lesssim 18$) close in crowded regions ($|b|\lesssim 5$ deg) un-corrected calibration effects become significant inflating the systematic uncertainties. When using \gaia DR2 astrometry to search for excess noise from genuine source characteristics, these systematic uncertainties should be taken into account.


\section{Unit Weight Error}
\label{sec:uwe}

\begin{figure}
  \centering
  \includegraphics[width=\columnwidth]{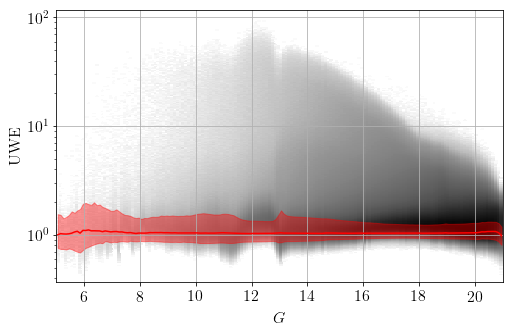}
  \caption[]{The reduced $\chi^2$ of the astrometric solution, UWE is estimated from the published covariances using the predicted covariance for simple point sources producing a distribution with median $\sim1$ (red solid). The distribution of source (black histogram, log normalised) extends out to high values of UWE due to sources with high excess noise.}
   \label{fig:uwemag}
\end{figure}
\begin{figure*}
  \centering
  \includegraphics[width=\textwidth]{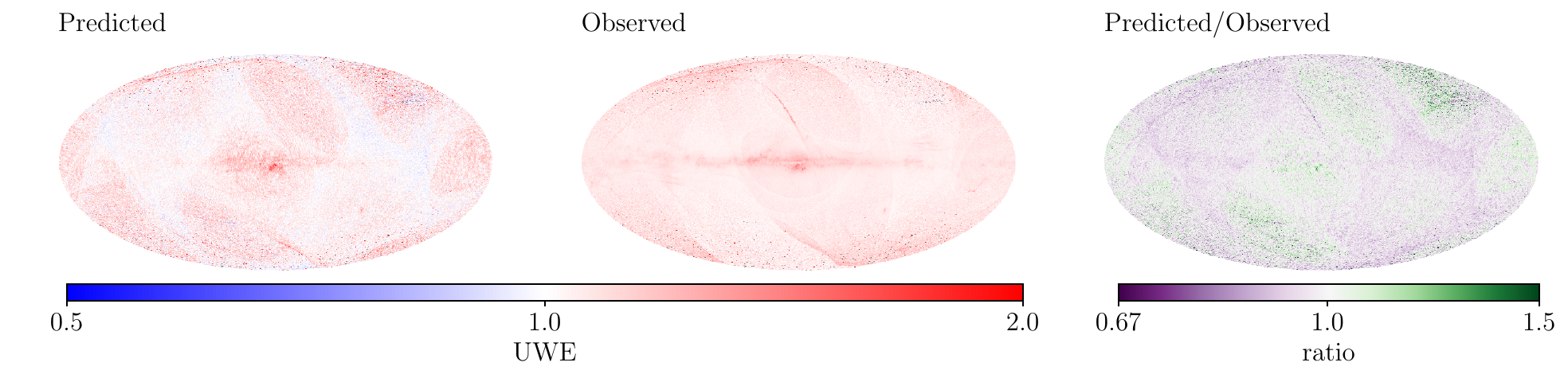}
  \caption[]{Predicted UWE for sources with $G\in[18.1,19.0]$ in Galactic coordinates (left) show little structure on the sky as expected. The median observed UWE in HEALPix level 7 bins (middle) has some limited structure relating to problematic individual scans in \gaia DR2. The ratio between predicted and observed UWE shows strong agreement down to the $~10\%$ level at which point the DoF bug will introduce a bias to the published values for $G\sim18$.}
   \label{fig:uwehpx}
\end{figure*}

Unit Weight Error (UWE) is the reduced chi-squared statistic of the astrometric fit to observations.
\begin{equation}
    \mathrm{UWE} = \sqrt{\frac{1}{\nu}(\mathbf{x}' - \mathbf{x})^\mathrm{T} \mathbfss{K}^{-1}(\mathbf{x}' - \mathbf{x})}
\end{equation}
where $\mathbf{x}'$ and $\mathbf{x}$ are the measured and expected position measurements of a source, $\mathbfss{K}=\mathrm{diag}[\sigma_1^2, \sigma_2^2...\sigma_N^2]$ is the measurement covariance and $\nu = N-5$ is the number of degrees of freedom. 

For simple point sources UWE will be drawn from a Gamma distribution, $\mathrm{UWE}\sim \Gamma\left[\nu/2, \nu/2\right]$ such that the expected value is 1 and the variance is inversely proportional to the degrees of freedom. However any excess stellar motion or an extended flux distribution will introduce an excess UWE above 1 as happens for binary systems \citep{Penoyre2020} or astrometric microlensing events \citep{McGill2020}. \gaia publishes $\chi^2$ and the degrees of freedom $\nu = N-5$
for all stars with 5D astrometry in DR2 from which UWE can be calculated. However, the published $\chi^2$ is plagued by the DoF bug \citepalias{Lindegren2018} which makes values unreliable to use for estimating the excess noise.

This can be remedied by renormalising the published UWE as a function of colour and apparent magnitude to produce a new statistic, RUWE\footnote{\url{http://www.rssd.esa.int/doc_fetch.php?id=3757412}}. RUWE is normalised such that the $41^\mathrm{st}$ percentile is 1 as this was found to represent well behaved sources where the median showed significant contamination from sources with excess error. This works well at face value and produces a usable statistic however there are two limitations. Firstly RUWE does not follow a well defined $\chi^2$ distribution as would be expected from UWE, therefore estimating the significance of excess noise is challenging. Secondly, in cases where excess noise is not equally likely in all colours and apparent magnitudes, the renormalisation can hide some of the expected excess. This would be problematic when establishing the binary fraction as a function of colour and absolute magnitude which is expected to vary considerably between stellar populations \citep{PriceW2020, Belokurov2020}.



An alternative of UWE for a source with measured 5D astrometry is given by
\begin{equation}
\mathrm{UWE} = \sqrt{\frac{1}{n}\mathbb{E}[\boldsymbol{\delta}^{\mathrm{T}}\Sigma^{-1} \boldsymbol{\delta}] }
\end{equation}
where $n=5$ is the dimensionality of the astrometry. Given $\boldsymbol{\delta}=(\mathbf{r'}-\mathbf{r})\sim \mathcal{N}(0,\mathbfss{C})$
\begin{align}
\mathbb{E}[\boldsymbol{\delta}^\mathrm{T}\Sigma^{-1} \boldsymbol{\delta}] &= \int \frac{1}{(2\pi)^{n/2}\sqrt{||\mathbfss{C}||}} (\boldsymbol{\delta}^\mathrm{T} \Sigma^{-1} \boldsymbol{\delta}) \exp \left( - \frac{1}{2} (\boldsymbol{\delta}^\mathrm{T} \mathbfss{C}^{-1} \boldsymbol{\delta}) \right)  \mathrm{d}\boldsymbol{\delta} \nonumber\\
&= \int \frac{1}{(2\pi)^{n/2}} (\mathbf{y}^\mathrm{T} \sqrt{\mathbfss{C}}^\mathrm{T} \Sigma^{-1} \sqrt{\mathbfss{C}} \mathbf{y}) \exp\left( - \frac{1}{2} (\mathbf{y}^\mathrm{T} \mathbf{y}) \right) \mathrm{d}\mathbf{y}
\end{align}
where $\mathbf{y} = \sqrt{\mathbfss{C}^{-1}} \boldsymbol{\delta}$ and $\frac{\mathrm{d}\mathbf{y}}{\mathrm{d}\boldsymbol{\delta}} = \sqrt{||\mathbfss{C}||}$. Letting $W =\sqrt{\mathbfss{C}}^\mathrm{T} \Sigma^{-1} \sqrt{\mathbfss{C}}$
\begin{equation}
\mathbb{E}[\boldsymbol{\delta}^\mathrm{T}\Sigma^{-1} \boldsymbol{\delta}] = \int \frac{1}{(2\pi)^{n/2}} (\mathbf{y}^\mathrm{T} \mathbfss{W} \mathbf{y}) \exp\left(- \frac{1}{2} (\mathbf{y}^\mathrm{T} \mathbf{y}) \right)  \mathrm{d}\mathbf{y}.
\end{equation}
All off-diagonal elements of $\mathbfss{W}$ produce antisymmetric integrands in $\mathbf{y}$ leaving only the diagonal elements
\begin{align}
\mathbb{E}[\boldsymbol{\delta}^\mathrm{T}\Sigma^{-1} \boldsymbol{\delta}] = & \sum_{i=1}^n \int \frac{1}{(2\pi)^{n/2}} \mathbfss{W}_{i,i} \mathbf{y}_i^2 \exp\left(- \frac{1}{2} \mathbf{y}_i^2 \right)  \mathrm{d}\mathbf{y} \nonumber\\
= & \sum_{i=1}^n \mathbfss{W}_{i,i}.
\end{align}
Substituting $\Sigma$ back into this we have UWE in terms of the published covariance $C$
\begin{align}
\mathrm{UWE} &= \sqrt{\frac{1}{5}\mathrm{Tr}(\sqrt{\mathbfss{C}} \Sigma^{-1} \sqrt{\mathbfss{C}})} \nonumber \\
&= \sqrt{\frac{1}{5}\mathrm{Tr}(\mathbfss{C} \Sigma^{-1})}.
\end{align}

Using this formula, we estimate UWE for all stars with 5D astrometry in \gaia DR2. The distribution of UWE as a function of magnitude, shown in the Fig.~\ref{fig:uwemag}, is uniform with the median $\langle\mathrm{UWE}\rangle\gtrsim 1$. The fact that the median UWE sits slightly higher than 1 is due to the contribution from sources with excess noise. The spread of UWE which is greatest at $G\sim13$ and narrows to fainter magnitude is a clear signature of excess error which is resolvable at brighter magnitude but becomes increasingly dominated by photon count noise for fainter sources. 

Our estimate is compared with the published UWE for sources with $G\in[18.1,19.0]$ in Fig.~\ref{fig:uwehpx}. At these dim magnitudes, the impact of the DoF bug is small. Across the sky, our estimate of UWE is in excellent agreement with the published value producing no systematic residual signal in the right hand panel down to $~10\%$ uncertainty.

In \gaia EDR3, the DoF bug is be fixed and our estimate of UWE will be superseded by the published value. However, the fact that our measurement is in good agreement with the published UWE is indicative that the published covariance alongside our prediction of the ASF contains all of the information contained in UWE and more. Whilst UWE can be used to determine the probability and amplitude of any excess variance, the ASF has the potential to decode the orientation and time variation of excess noise.




\section{Astrometric Selection}
\label{sec:sigma5dmax}

\begin{figure}
  \centering
  \includegraphics[width=\columnwidth]{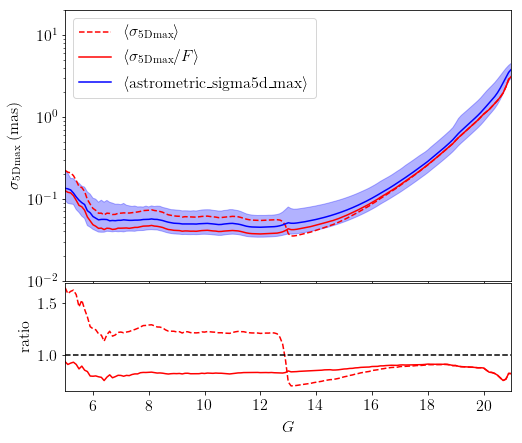}
  \caption[]{Predicted $\sigma_\mathrm{5Dmax}$ after correcting for the DoF bug (red solid) as a function of magnitude for all sources in the \gaia DR2 astrometry shows strong agreement with the published values (median - blue solid, $16^\mathrm{th}-84^\mathrm{th}$ percentiles - blue shaded). The model before correcting for the DoF bug (red dashed) shifts at $G\sim13$ the magnitude at \gaia switches from 2D to 1D observations. The systematic underestimate of the prediction against the median published astrometry is expected to be due to remaining calibration uncertainties which we have not fully accounted for.}
   \label{fig:sigma5d_max_mag}
\end{figure}
\begin{figure*}
  \centering
  \includegraphics[width=\textwidth]{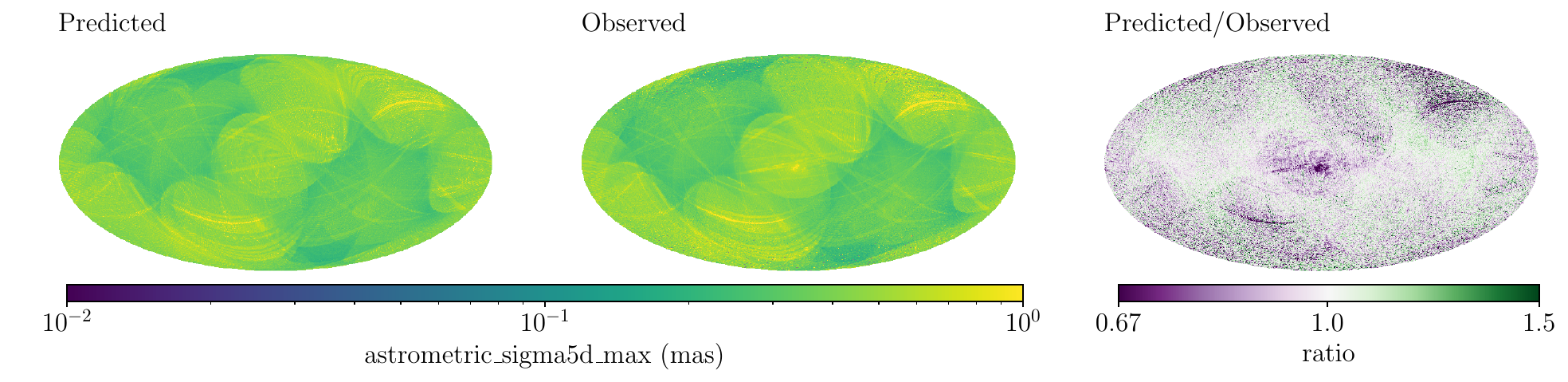}
  \caption[]{The predicted $\sigma_\mathrm{5Dmax}$ (left) for $G\in[18.1,19.0]$ agrees well with the HEALPix level 7 binned median published values (middle) across the sky in Galactic coordinates. The ratio between the predicted and observed shows some weak residuals in the Galactic bulge and a bad scan which impacts areas of sky which haven't been significantly observed in \gaia DR2.}
   \label{fig:sigma5d_max_hpx}
\end{figure*}
\begin{figure}
  \centering
  \includegraphics[width=\columnwidth]{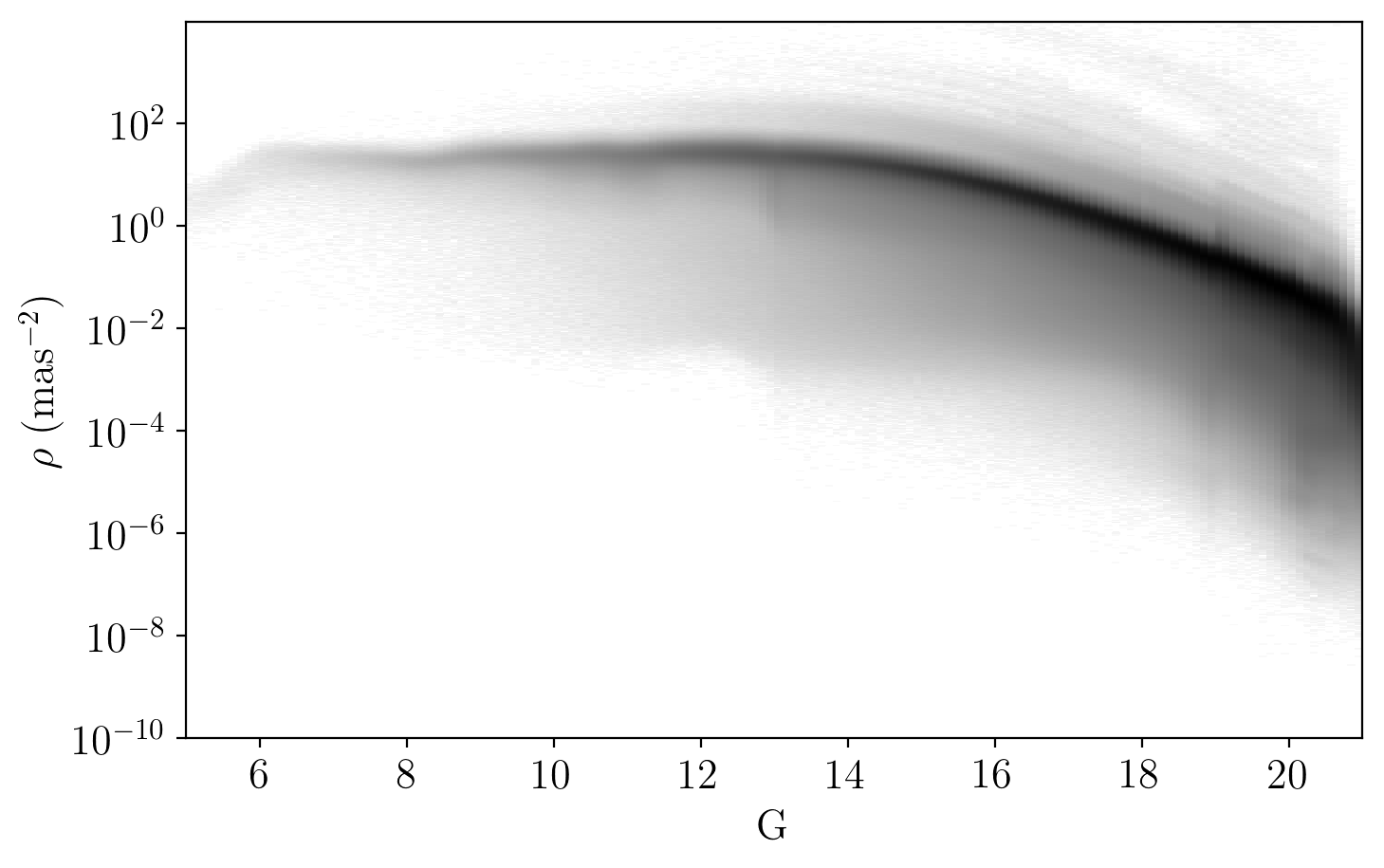}
  \caption[]{$\rho$ encodes the magnitude dependence of the predicted astrometric precision of \gaia DR2. 5D astrometric covariance is only published for the subset of DR2 with 5D astrometry however $\sigma_\mathrm{5Dmax}$ is published for all sources in DR2. We estimate $\rho$ for all sources in DR2 with $k_\mathrm{VP}>5$ using Eq.~\ref{eq:rho_sigma} shown here as a function of magnitude.}
   \label{fig:rho}
\end{figure}
\begin{figure*}
  \centering
  \includegraphics[width=\textwidth]{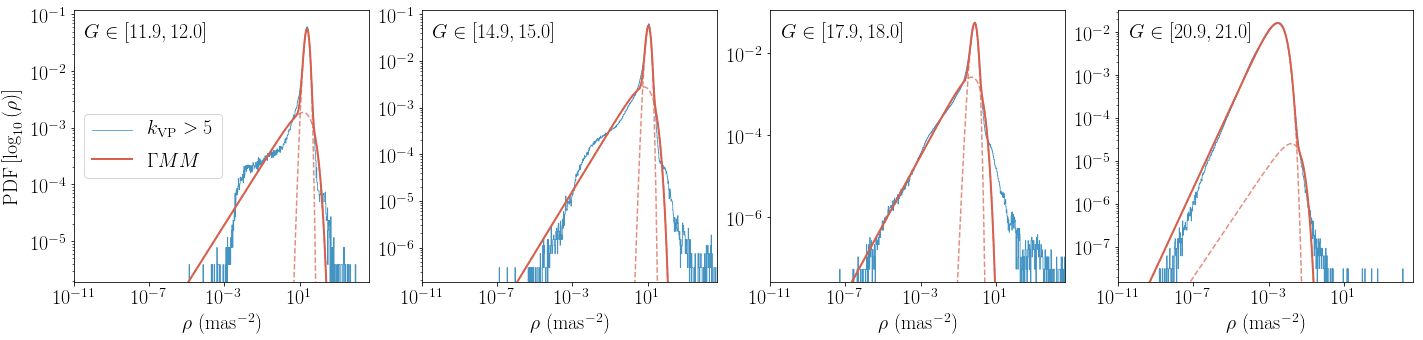}
  \caption[]{The distribution of $\rho$ in $0.1$mag bins for \gaia DR2 sources with $k_\mathrm{VP}>5$ (blue histograms) consists of a sharp peak of well behaved inertial point sources with a long wing to low $\rho$ from sources with high excess error. This is fit with a two component Gamma Mixture Model ($\Gamma$MM) with one component fitting the peak and the second accounting for the low-$\rho$ wing (red solid line). Red-dashed lines show the individual $\Gamma$ components.}
   \label{fig:rhodist}
\end{figure*}
\begin{figure}
  \centering
  \includegraphics[width=\columnwidth]{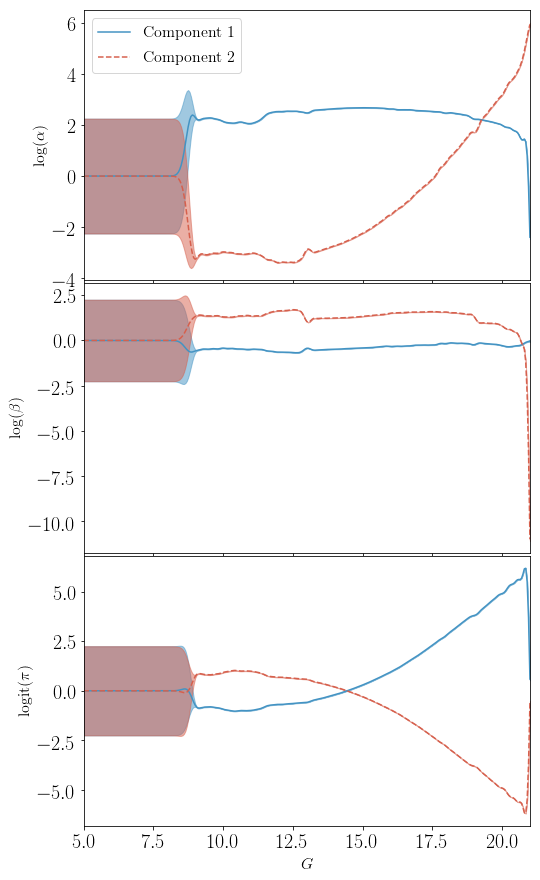}
  \caption[]{The five parameters of the two component $\Gamma$MM are fit with with a single $\mathcal{GP}$ as a function of magnitude with a square exponential covariance kernel for matching parameters using the posterior MCMC samples from each magnitude bin. Small volumes of data at the bright end mean that the $\mathcal{GP}$ is dominated by the prior with mean 0 and variance $s=2.578$.}
   \label{fig:rhoGP}
\end{figure}
\begin{figure*}
  \centering
  \includegraphics[width=\textwidth]{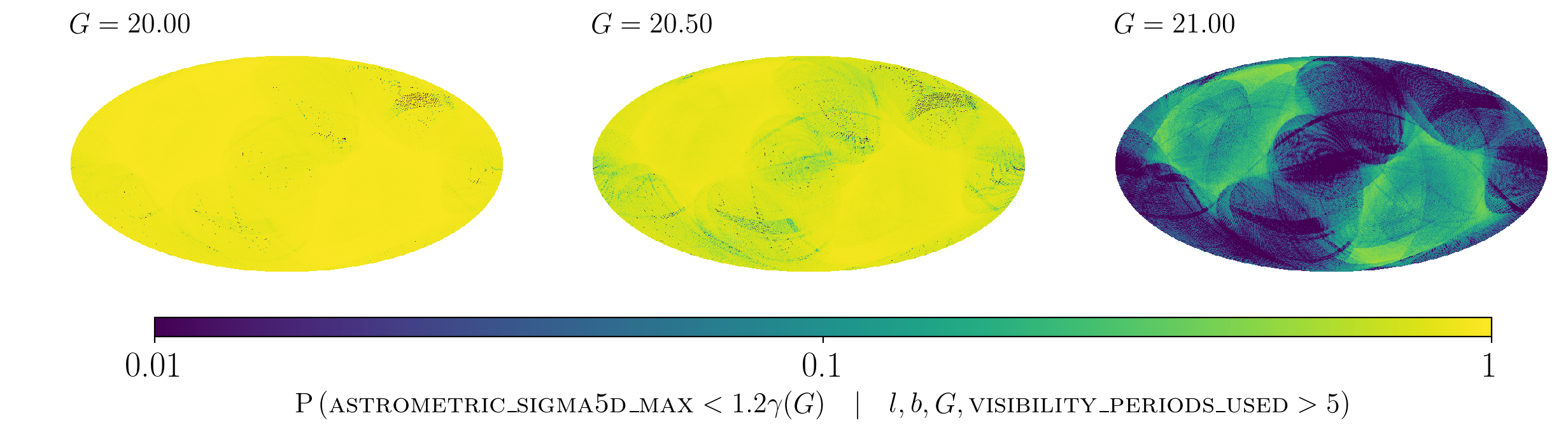}
  \caption[]{The selection probability of passing the $\sigma_\mathrm{5Dmax}$ cut is estimated from the $\Gamma$MM fits as a function of magnitude and position on the sky in HEALPix level 7 bins presented in Galactic coordinates. For $G\sim20$ (left), a negligible portion of stars are removed by the cut however at $G\sim21$ (right), the faintest magnitude for \gaia astrometry, the vast majority of sources in low scanned regions of the sky are removed from the sample.}
   \label{fig:sfprob}
\end{figure*}

In order to construct unbiased dynamical models of the Milky Way, it is critically important that we have a strong understanding of the completeness of our sample. \citet{Paper2} produced the selection function for the full \gaia DR2 catalogue however the subset of DR2 with 5D astrometry constitutes a biased subsample and therefore an astrometry selection function is required for studies which rely on parallax or proper motion data. The \gaia DR2 5D astrometry sample is the subset of the full sample that satisfies the cuts \citepalias[][Section 4.3]{Lindegren2018}:
\begin{itemize}
    \item $G<21$
    \item $\textsc{visibility\_periods\_used}>5$
    \item $\textsc{astrometric\_sigma5d\_max}>1.2\times\gamma(G)$
\end{itemize}
where $\gamma(G) = \mathrm{max}\left[1, 10^{0.2(G-18)}\right]$.

To construct the selection function for the 5D astrometry sample we can combine the effect of these cuts with the full sample selection function
\begin{equation}
    \mathrm{P}(\mathrm{S}_\mathrm{5D ast}) = \mathrm{P}( \mathrm{S}_\mathrm{5D ast} |  \mathrm{S}_\mathrm{DR2})  \mathrm{P}(\mathrm{S}_\mathrm{DR2}) 
\end{equation}
where $\mathrm{S}_\mathrm{5D ast}$ is the event that a source is published with 5D astrometry and $\mathrm{S}_\mathrm{DR2}$ is the event that a source is included in DR2 with or without 5D astrometry. $\mathrm{P}(\mathrm{S}_\mathrm{DR2})$ is the full \gaia DR2 selection function estimated in \citet{Paper2}. The probability of a star in DR2 receiving 5D astrometry, $\mathrm{P}( \mathrm{S}_\mathrm{5D ast} |  \mathrm{S}_\mathrm{DR2})$, is governed by the three cuts outlined above.

The second cut on \textsc{visibility\_periods\_used} ($k_\mathrm{VP}$) is a complex function of the scanning law and detection probability and will be the subject of a future work. Here we will focus on the \textsc{astrometric\_sigma5d\_max} ($\sigma_\mathrm{5Dmax}$) cut.

$\sigma_\mathrm{5Dmax}^2$ is the maximum eigenvalue of the scaled astrometric covariance matrix 
\begin{equation}
\sigma_\mathrm{5Dmax}^2 = \lambda_\mathrm{max}\left[S\,C\,S\right]
\end{equation}
where $C \in \mathbb{R}^{5\times5}$ is the published 5D covariance matrix and ${{S = \mathrm{diag}[1,1,\sin{(\xi)},T/2,T/2]}}$
where $\xi=45\;\mathrm{deg}$ is the solar aspect angle of the \gaia satellite and $T=1.75115\;\mathrm{yr}$ is the time window of observations used in \gaia DR2.

Our aim here is to estimate the contribution to the selection function solely from the cut on $\sigma_\mathrm{5Dmax}$, 
\begin{equation}
    \mathrm{P}(\sigma_\mathrm{5Dmax}<1.2\gamma \,|\,k_\mathrm{VP}>5, G, l, b).
\end{equation}
$\sigma_\mathrm{5Dmax}$ and $k_\mathrm{VP}$ are published for all sources in \gaia DR2 so this could be easily achieved by taking the ratio of number of sources with $\sigma_\mathrm{5Dmax}<1.2\gamma(G)$ and $k_\mathrm{VP}>5$ to only those with $k_\mathrm{VP}>5$ as a function of apparent magnitude and position on the sky
\begin{align}
    \mathrm{P}(\sigma_\mathrm{5Dmax}<1.2\gamma \,&|\,k_\mathrm{VP}>5, G,l,b) \nonumber\\ &= \frac{N(\sigma_\mathrm{5Dmax}<1.2\gamma,\, k_\mathrm{VP}>5, G, l, b)}{N(k_\mathrm{VP}>5, G, l, b)}.
\end{align}
This approach is limited by Poisson count noise. To resolve scanning law variations, one would need to resolve the sky to at least HEALPix level 7. Using 200 magnitude bins, this results in an average of $\sim30$ stars with astrometry per bin which will be dominated by the Milky Way disk. At high latitudes the inference will be entirely dominated by Poisson noise.

Instead, we can use \gaia's predicted covariance as a function of position on the sky given in Section~\ref{sec:method}. This enables us to reach unlimited resolution on the sky without HEALPix binning the data. We can predict $\sigma_\mathrm{5Dmax}$ for any source in \gaia as a function of magnitude and position on the sky
\begin{align}
    \sigma_\mathrm{5Dmax} &= \sqrt{\lambda_\mathrm{max}\left[ S \Sigma S \right]}\\
    & = \frac{1}{\sqrt{\rho(G)}} \sqrt{\lambda_\mathrm{max}\left[ S \Phi^{-1} S \right]}.
\end{align}
where we have used the substitution $\Sigma^{-1}=\rho(G)\Phi$ from Eq.~\ref{eq:final_prec} and $\rho(G)$ is defined in Eq.~\ref{eq:rhodef}. A comparison of the running median of the predicted $\sigma_\mathrm{5Dmax}$ (red dashed) and observed \textsc{astrometric\_sigma5d\_max} (blue solid) in Fig.~\ref{fig:sigma5d_max_mag} shows that the prediction overestimates for $G<13$ and underestimates for $13<G<16$. The cause of this is the `DoF' bug detailed in \citetalias[Appendix A]{Lindegren2018}. Our predicted $\sigma_\mathrm{5Dmax}$ has been corrected for the DoF bug whilst the published values, on which the astrometry was selected, had not been corrected. The DoF bug is de-corrected from our prediction dividing through by a factor $F$ from Eq.~\ref{eq:Fdof} to produce the red solid line, in good agreement with the published $\sigma_\mathrm{5Dmax}$ as a function of magnitude. The predicted value marginally systematically underestimates $\sigma_\mathrm{5Dmax}$ across all magnitudes by $\sim10\%$ which we conjecture may be linked to time dependence of $\sigma_\mathrm{AL}$ which produces systematic uncertainties at the same level however the exact cause of this discrepancy for $\sigma_\mathrm{5Dmax}$ is unclear.

The predicted and observed distribution of $\sigma_\mathrm{5Dmax}$ on the sky are shown in Fig.~\ref{fig:sigma5d_max_hpx} with the right panel showing strong agreement across the majority of the sky. Some residual streaks still persist in the South East and North West regions of the sky which match those seen in Section~\ref{sec:results} when comparing the predicted and observed astrometry variances. These correspond to broken scans in \gaia DR2 which haven't previously been diagnosed. We use the HEALPix time extractor tool \citep{Paper5} to constrain the times at which these scans happened in DR2. The clearest time ranges are given in Table~\ref{tab:badscans} where the time range OBMT$=1556-1560$rev is the direct cause of the residual streaks discussed above.

\begin{table}
\centering
\caption{Time periods producing un-modelled scan features in \textsc{astrometric\_sigma5d\_max}. All times are given in OBMT (rev).}
\label{tab:badscans}
\setlength\tabcolsep{10pt}
\begin{tabular}{lll}
\hline Start & End  & Magnitudes  \\ 
\hline
1447  &  1449  &   19.05 - 19.95   \\
1453  &  1457  &   20.00 - 21.00   \\
1556  &  1560  &   18.10 - 21.00   \\
1730  &  1732  &   20.00 - 21.00   \\
\hline
\end{tabular}
\end{table}

$\sigma_\mathrm{5Dmax}$ is published for all sources in \gaia DR2 whether or not they have published 5D astrometry. We can therefore use the published $\sigma_\mathrm{5Dmax}$ to estimate $\rho$ for all stars in DR2
\begin{equation}
    \rho = \frac{\lambda_\mathrm{max}\left[S\Phi^{-1}S\right]}{\sigma_\mathrm{5Dmax}^2}.
    \label{eq:rho_sigma}
\end{equation}
The distribution of $\rho$ as a function of magnitude is shown in Fig.~\ref{fig:rho} where the distribution is largely flat at brighter magnitudes whilst declining for $G>13$ due to low photon count noise. The spread to lower values is driven by excess noise due to binaries and other accelerating or extended sources. 

In every $0.1$ mag bin we fit a two component Gamma mixture model ($\Gamma$MM) to model the distribution of $\rho$,
\begin{equation}
    \mathrm{P}(\rho) = \pi_1\Gamma(\rho ; \alpha_1,\beta_1) + \pi_2\Gamma(\rho ; \alpha_2,\beta_2).
\end{equation}
One component of the mixture model fits the peak of the distribution which is dominated by well behaved simple point sources whilst the second component has an extended tail to low $\rho$ which accounts for sources with significant excess noise. Examples of these fits in four magnitude bins are shown in Fig.~\ref{fig:rhodist} demonstrating reasonable agreement at dim magnitudes whilst somewhat cutting through the low $\rho$ tail at bright magnitudes. At dim magnitudes, there is also a small excess of sources at large $\rho$. The precise cause of this tail is unclear but since any cuts on $\sigma_\mathrm{5Dmax}$ will be on the low $\rho$ end, the fact that we haven't correctly modelled the high $\rho$ tail will only generate a $<1\%$ systematic uncertainty in the inferred selection function. Priors used for each of the parameters in the $\Gamma$MM are given in Table~\ref{tab:GMMpriors}. The parameters are fit using expectation maximisation and posterior distributions produced using emcee \citep{Foreman2013}. 

The behaviour of the $\Gamma$MM parameters as a function of magnitude is modelled with a single Gaussian Process. For values of the same parameter at different magnitudes, the $\mathcal{GP}$ uses a square exponential kernel with variance $s$ and scale length $l$. For different parameters, we assume no intrinsic correlation, however, correlations will be introduced between different parameters of the same magnitude bin through the covariance of MCMC samples. Applying $k$-fold cross validation with $k=5$ we infer hyperparameter values of $l=0.224$, $s=2.578$. The posterior $\mathcal{GP}$ is shown in Fig.~\ref{fig:rhoGP} where the blue solid and red dashed lines are the two components for each parameter. Due to a lack of bright sources in \gaia DR2 astrometry, the $\mathcal{GP}$ at the bright end is dominated by the prior from the kernel. Since a negligible proportion of stars will be influenced by the $\sigma_\mathrm{5Dmax}$ cut at these magnitudes, this is not a significant issue for the model.

Using the $\Gamma$MM as a function of magnitude, the selection function probability is given by
\begin{align}
    \mathrm{P}(\sigma_\mathrm{5Dmax}<1.2\gamma(G) &| k_\mathrm{VP}>5) \nonumber\\&= \int_{\rho_\mathrm{min}}^\infty \sum_{j=1}^2 \pi_j(G)\Gamma\left[\rho; \alpha_j(G), \beta_j(G)\right] \mathrm{d}\rho
\end{align}
where $\rho_\mathrm{min} = \frac{\lambda_\mathrm{max}\left[S\Phi S\right]}{(1.2\gamma(G))^2}$ from substituting $\sigma_\mathrm{5Dmax}=1.2\gamma(G)$ into Eq.~\ref{eq:rho_sigma}.

The selection probability is given at three magnitudes in Fig.~\ref{fig:sfprob} demonstrating that the cut only has a significant effect for $G>20$. At the faintest magnitudes, regions of the sky which have been only sparsely scanned in \gaia DR2 are most likely to be removed due to the cut on $\sigma_\mathrm{5Dmax}$. In the most extreme cases such as in the Milky Way bulge, this can result in $<1\%$ completeness in the \gaia DR2 astrometry sample.

\begin{table}
\centering
\begin{tabular}{|| c | c || }
\hline
$\log(\alpha)$ & $\mathrm{U}[-\infty, \infty]$\\
$\log(\beta)$  & $\mathrm{U}[-\infty, \infty]$\\
$\pi$ &  $\mathrm{Dirichlet}(a=[2,2])$\\
\hline
\end{tabular}
\caption{Priors used for $\Gamma$MM fit to $\rho$ distribution.}
\label{tab:GMMpriors}
\end{table}

Due to the simplicity of our 2 component $\Gamma\mathrm{MM}$, the fits to the distribution of stars can produce significant offsets from the true distribution of data at the low $\rho$ tail as is seen in the fourth panel of Fig.~\ref{fig:rhodist}. The overestimate of the number of sources at low $\rho$ in this case
will lead to a significant overestimate of the number of stars with high $\sigma_\mathrm{5Dmax}$ which will subsequently get cut from the 5D astrometry sample. For this work we consider the method a proof of principle for applying the ASF in order to derive the selection function and will refine the fits to $\rho$ as a function of magnitude when producing the full \gaia DR2 5D astrometry selection function.

\section{Discussion}
\label{sec:discussion}

\subsection{Excess Covariance}

In this work we have derived and discussed the importance of the ASF for analysing simple point sources in \gaia DR2. However we haven't established how to use the ASF to estimate the excess covariance or precisely how this can be interpreted.

Consider a source with true 5D astrometry, $r$. However the source is not a simple point source such that the apparent position as a function of time is not well modelled by the 5D astrometric solution. If the excess noise may be parameterised by a 5D covariance, E, the probability of measuring the apparent 5D astrometry as $\mathbf{r}_E$ will be given by
\begin{equation}
    \prob(\mathbf{r}_E) = \mathcal{N} ( \mathbf{r}_E \,;\, \mathbf{r}, \mathbfss{E}).
\end{equation}
If one attempts to measure this source, the uncertainty with which the 5D astrometry is measured is given by the ASF
\begin{equation}
    \prob(\mathbf{r}') = \mathcal{N} ( \mathbf{r}' \,;\, \mathbf{r}_E, \Sigma).
\end{equation}
By multiplying the two distributions together and marginalising over $\mathbf{r}_E$, we can determine the probability distribution of the measured 5D astrometry
\begin{align}
    \prob(\mathbf{r}') &= \int \mathrm{d}^5\mathbf{r}_E \,\mathcal{N} ( \mathbf{r}' \,;\, \mathbf{r}_E, \Sigma)\,\mathcal{N} ( \mathbf{r}_E \,;\, \mathbf{r}, \mathbfss{E})\\
    &=\mathcal{N} \left( \mathbf{r}' \,;\, \mathbf{r}, \left(\Sigma^{-1}+\mathbfss{E}^{-1}\right)^{-1}\right)\nonumber\\
    &=\mathcal{N} ( \mathbf{r}' \,;\, \mathbf{r}, \mathbfss{C}).\nonumber
\end{align}

Therefore, in this vastly oversimplified situation, the final measurement uncertainty for the 5D astrometry is given by the convolution of the excess noise and the ASF (providing the contribution from the observation measurement uncertainty). 

There are two significant issues with this interpretation when considering the astrometry published by \gaia. Firstly, the AGIS pipeline does not formally infer the measurement uncertainty induced by excess noise. Residuals beyond simple point source astrometry are absorbed into a 1D excess noise parameter for each source as well as impacting the weights used for the given observations. The second problem is that source excess noise can disguise itself as a shift in the simple point source astrometry. As shown in \citet{Penoyre2020}, excess binary motion can have complex effects on the posterior astrometry from \gaia including a phenomenon called the proper motion anomaly \citep{Kervella2019}. Interpretation of the excess covariance will require simulating stellar populations and emulating the AGIS pipeline in order to forward model how the intrinsic properties of the source relate to the posterior excess.




\subsection{Mock observations}

Whilst we have entirely focused on the implications of the ASF for constraining excess source noise, it is also directly applicable to simulations in order to generate mock \gaia catalogues for Milky Way analogues.

Recent simulations such as Auriga \citep{Grand2017} and VINTERGATAN \citep{Agertz2020} have demonstrated the ability of the latest generation of cosmological simulations to produce Milky Way analogues which are excellent tools for studying the physical processes which govern the evolution of our galaxy. Performing a direct comparison with \gaia observations requires the \gaia selection functions and measurement uncertainty. The ASF provides the expected uncertainty of 5D astrometry for a simple point source. Given a simulated star with astrometry $\mathbf{r}$ as observed from the sun, the astrometry that would be measured by \gaia, $\mathbf{r}'$ can be inferred by sampling from the ASF
\begin{equation}
    \mathbf{r}' \sim \mathcal{N}(\mathbf{r}, \Sigma(G, l, b)).
\end{equation}




\section{Accessing the ASF}
\label{sec:code}

The ASF is a useful tool for inferring excess astrometric covariance of \gaia 5D astrometry sources. To make this accessible, we've added a module to the \textsc{Python} package \textsc{scanninglaw} (\url{https://github.com/gaiaverse/scanninglaw}) \citep{Paper3}. The user can ask the question `What astrometric covariance would \gaia have published if my star was a simple point source?'.

As always, this is demonstrated by determining the ASF covariance of the fastest main-sequence star in the Galaxy \citep[S5-HVS1,][]{Koposov2020} for \gaia DR2. The diagonal elements of the output covariance give the variance in $\alpha_0^*$, $\delta_0$, $\varpi$ ($\mathrm{mas}^2$), $\mu_\alpha^*$, $\mu_\delta^*$ ($\mathrm{mas}^2/\mathrm{y}^2$).
\begin{lstlisting}[language=Python]
import scanninglaw.asf as asf
from scanninglaw.source import Source

dr2_sl = asf.dr2_asf(version='cog')
s5_hvs1 = Source('22h54m51.68s',
                 '-51d11m44.19s',
           photometry={'gaia_g':16.02},
           frame='icrs')
Sigma = dr2_sl(s5_hvs1)

print('ASF Covariance: \n', Sigma)

>> ASF Covariance Position: 
[[ 0.0005,  0.0004, -0.0005,  0.0003,  0.0006],
 [ 0.0004,  0.0029, -0.0023,  0.0016,  0.0013],
 [-0.0005, -0.0023,  0.0057, -0.0026, -0.0034],
 [ 0.0003,  0.0016, -0.0026,  0.0038,  0.0017],
 [ 0.0006,  0.0013, -0.0034,  0.0017,  0.0096]]
\end{lstlisting}

\section{Conclusion}

The Astrometry Spread Function is the astrometric uncertainty distribution which would be  expected for a point source with linear motion relative to the solar system barycenter (simple point source) given the source apparent magnitude and position on the sky. \gaia's DPAC estimate the astrometric solution using an iterative linear regression algorithm. Given the uncertainty of individual observations and the scanning law, we have been able to reconstruct the astrometric covariance that would be expected for a simple point source observed by \gaia DR2. The ASF is a 5D multivariate Gaussian distribution with mean $\mathbf{0}$ and covariance $\Sigma \in \mathbb{R}^{5\times5}$ where we have formally derived $\Sigma(G,l,b)$.

Assuming the bulk of stars in the \gaia DR2 5D astrometry sample are simple point sources down to \gaia's detection limit, we compare our result with the published covariances and find extremely good agreement down to sub-degree scales on the sky. The only region with marginal disagreement is the highest source density regions of the bulge where the combination of source crowding and few scans in \gaia DR2 invalidate our assumptions. Therefore we caution the use of the ASF in highly crowded regions with low scan counts. 

We used the ASF in combination with the published covariance to infer unit weight error for \gaia DR2 sources. The strong agreement with the published UWE demonstrates that the ASF can be used to find the excess error in \gaia observations due to physical source characteristics. The ASF will be a valuable tool for exploiting \gaia data to model binary stars, astrometric microlens events and extended sources.

Finally we applied the ASF to predict the selection function contribution from the cut on \textsc{astrometric\_sigma5d\_max} used to generate the \gaia DR2 5D astrometry sample. This will be a key component of the full astrometry selection function which is a vital tool for unbiased modelling of Milky Way kinematics from \gaia's 5D astrometry.

\section*{Acknowledgements}
AE thanks the Science and Technology Facilities Council of
the United Kingdom for financial support. DB thanks Magdalen College for his fellowship and the Rudolf Peierls Centre for Theoretical Physics for providing office space and travel funds. This work has made use of data from the European Space Agency (ESA) mission
{\it Gaia} (\url{https://www.cosmos.esa.int/gaia}), processed by the {\it Gaia}
Data Processing and Analysis Consortium (DPAC,
\url{https://www.cosmos.esa.int/web/gaia/dpac/consortium}). Funding for the DPAC
has been provided by national institutions, in particular the institutions
participating in the {\it Gaia} Multilateral Agreement.

\section*{Data availability}
The data underlying this article are publicly available from the European Space Agency's \gaia archive (\url{https://gea.esac.esa.int/archive/}). 
The ASF covariance matrix derived in Section~\ref{sec:method} is publicly available on the Harvard Dataverse (\url{https://doi.org/10.7910/DVN/FURYBN}) and accessible through the \textsc{GitHub} repository \textsc{scanninglaw} (\url{https://github.com/gaiaverse/scanninglaw}).




\bibliographystyle{mnras}
\bibliography{references} 




\appendix

\section{Time dependence}
\label{sec:timedep}

\begin{figure}
  \centering
  \includegraphics[width=0.495\textwidth]{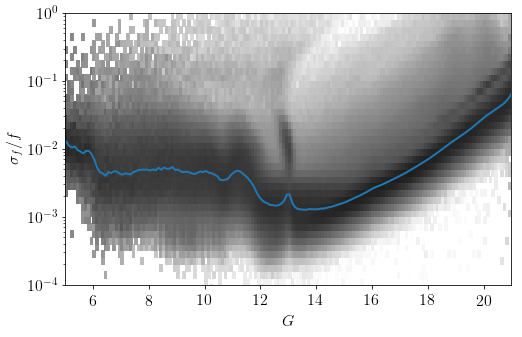}
  \caption[]{The relative flux error as a function of magnitude for all observations in the \gaia epoch photometry\citep{Riello2018, Evans2018, Holl2018} (black histograms, log normalised) shows complex structure due to changes in window class configurations\citep{Riello2018}. The running median (blue solid line) shows similar structure to $\sigma_\mathrm{AL}$ in Fig.~\ref{fig:sigma_al} with both dominated by photon count noise.}
   \label{fig:sigmaf_mag}
\end{figure}
\begin{figure}
  \centering
  \includegraphics[width=0.495\textwidth]{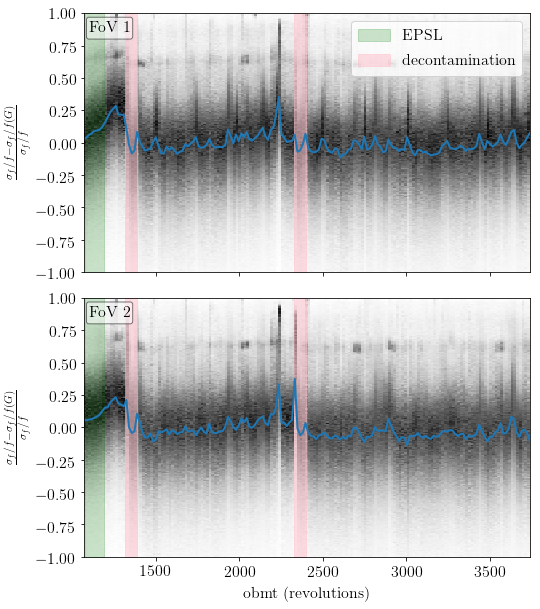}
  \caption[]{The relative flux error, recentered and renormalised by the median as a function of magnitude from Fig.~\ref{fig:sigmaf_mag}, varies as a function of time throughout \gaia DR2 due to mirror contamination, micro-meteoroid impacts and variations in background flux. Data taken during the Ecliptic Polar Scanning Law (green) and decontamination events (pink) we not used in the \gaia DR2 astrometry. The running median (blue solid) of the observations does not vary significantly from zero relative to the variance in individual measurements. This justifies the assumption that the astrometric measurement uncertainty is not a strong function of time.}
   \label{fig:sigmaf_time}
\end{figure}

To estimate the \gaia astrometric precision, we have assumed that all observations at the same apparent magnitude have the same precision. This assumption will break down if the precision of \gaia is time dependent. 

Without any epoch astrometry, it is challenging to assess the scale of the impact of this time dependence on the posterior precision. We are however provided epoch photometry for 550,737 sources. The astrometric uncertainty should scale with $\sigma_\mathrm{AL} \sim \sigma_f/f$ where $f$ is the observed flux of the source as both are dominated by photon count noise. The centroid is actually mainly sensitive to the slope of the wings of the LSF/PSF while the flux measurement to the core (as there is the most signal), but to first degree this relation should hold to understand how the centroid uncertainty depends on magnitude and time.

We take all 17,712,391 observations in the epoch photometry and find the median $\sigma_f/f$ in 0.2mag bins shown in Fig.~\ref{fig:sigmaf_mag}. We subtract the median off all data leaving the residuals. The distribution of residual errors against observation time is given in Fig.~\ref{fig:sigmaf_time} for each \gaia field-of-view (FoV). 
As \gaia operates, material from the satellite condenses and accumulates on the mirrors scattering light and reducing the precision of observations. To mitigate this, the spacecraft was heated up to evaporate the condensation and clean the mirrors \citep[see Section 4.2.1][]{Prusti2016}. These decontamination events (pink shaded regions) have taken place twice in Gaia DR2 \citepalias{Lindegren2018}.
The impact of the first decontamination event on the flux error is significant however at later times, the measurement precision does not degrade appreciably. In fact, any longer term trends are insignificant compared to the short term fluctuations on short ($\sim10$ revolution) timescales. 

The epoch flux supports the conjecture that the measurement precision of Gaia does not significantly change with time.

\section{DoF Bug}
\label{app:dof}

During the calibration of excess noise in \gaia DR2, the degrees of freedom parameter was erroneously used as the total number of AL and AC observations rather than only AL observations as intended. For a full explanation we refer the interested reader to \citetalias[Appendix A]{Lindegren2018}.

The result of this bug was that sources brighter than ${{G=13}}$, for which 2D observations were used, received overestimated measurement uncertainties. Through the attitude calibration, this indirectly impacted sources with ${{G>13}}$ although the increased photon count noise at dimmer magnitudes dampens the effect for dim stars.

To correct for this, the astrometric covariance was multiplied by a correction factor
\begin{equation}
F = (1+0.8R)\sqrt{\frac{2}{1+\sqrt{1+4(1+0.8R)^2\left(\frac{0.025\mathrm{mas}}{\sigma_\varpi}\right)}}}
\label{eq:Fdof}
\end{equation}
taken from \citetalias[Equation A.6]{Lindegren2018}. The published $\chi^2$ and \textsc{astrometric\_sigma5d\_max} didn't receive this correction, the latter being because the selection of the DR2 astrometry sample was performed before the bug was corrected.

As a result, when estimating \textsc{astrometric\_sigma5d\_max} in Section~\ref{sec:sigma5dmax}, in order to obtain a good agreement with the data, we must decorrect for the DoF bug by dividing through by the correction factor, $F$.

\bsp	
\label{lastpage}
\end{document}